\documentclass[fleqn,usenatbib]{mnras}

\usepackage{newtxtext,newtxmath}

\usepackage[T1]{fontenc}

\DeclareRobustCommand{\VAN}[3]{#2}
\let\VANthebibliography\thebibliography
\def\thebibliography{\DeclareRobustCommand{\VAN}[3]{##3}\VANthebibliography}

\usepackage{graphicx}	
\usepackage{amsmath}	
\usepackage{float}
\usepackage{pdflscape}
\usepackage{siunitx}
\usepackage{subcaption}
\usepackage{array}
\usepackage{rotating}
\usepackage{mathtools}
\usepackage{multicol}
\usepackage{multirow}
\usepackage{adjustbox}
\usepackage{nccmath}

\title[SN\,2020cpg: an energetic SE-SNe]{SN\,2020cpg: an energetic link between type IIb and Ib supernovae}

\author[Medler et al.]{K.~Medler$^{1}$ \thanks{E-mail: K.Medler@2019.ljmu.ac.uk},
P.~A.~Mazzali$^{1,2}$,
J.~Teffs $^{1}$,
S.~J.~Prentice $^{3}$,
C.~Ashall $^{4}$,
\newauthor
M.~Amenouche $^{5}$,
J.~P.~Anderson $^{6}$,
J.~Burke $^{7, 8}$,
T.~W.~Chen $^{9}$,
L.~Galbany $^{10}$,
\newauthor
M.~Gromadzki $^{11}$,
C.~P.~Guti\'errez $^{12, 13}$
D.~Hiramatsu $^{7, 8}$,
D.~A.~Howell $^{7, 8}$,
C.~Inserra $^{14}$,
\newauthor
E.~Kankare $^{13}$,
C.~McCully $^{7}$,
T.~E.~M\"uller-Bravo $^{15}$,
M.~Nicholl $^{16}$,
C.~Pellegrino $^{7, 8}$,
\newauthor
J.~Sollerman $^{9}$
\\
\\
$^{1}$Astrophysical Research Institute Liverpool John Moores University, Liverpool L3 5RF, UK \\
$^{2}$Max-Planck Institute for Astrophysics, Karl-Schwarzschild-Str. 1, D-85748 Garching, Germany \\
$^{3}$School of Physics, Trinity College Dublin, The University of Dublin, Dublin 2, Ireland\\
$^{4}$Institute for Astronomy, University of Hawai'i at Manoa, 2680 Woodlawn Dr., Hawai'i, HI 96822, USA \\
$^{5}$Universit\'{e} Clermont Auvergne, CNRS/IN2P3, LPC, Clermont-Ferrand, France \\
$^{6}$European Southern Observatory, Alonso de C\'ordova 3107, Casilla 19, Santiago, Chile \\
$^{7}$Las Cumbres Observatory Global Telescope Network, 6740 Cortona Dr. Suite 102, Goleta, CA 93117 \\
$^{8}$Department of Physics, University of California, Santa Barbara, USA \\
$^{9}$The Oskar Klein Centre, Department of Astronomy, Stockholm University, AlbaNova, SE-10691 Stockholm, Sweden \\
$^{10}$Departamento de F\'isica Te\'orica y del Cosmos, Universidad de Granada, E-18071 Granada, Spain \\
$^{11}$Astronomical Observatory, University of Warsaw, Al. Ujazdowskie 4, 00-478 Warszawa, Poland \\
$^{12}$Finnish Centre for Astronomy with ESO (FINCA), FI-20014 University of Turku, Finland \\ 
$^{13}$Tuorla Observatory, Department of Physics and Astronomy, FI-20014 University of Turku, Finland \\
$^{14}$School of Physics \& Astronomy, Cardiff University, Queens Buildings, The Parade, Cardiff, CF24 3AA, UK \\
$^{15}$School of Physics and Astronomy, University of Southampton, Southampton, Hampshire, SO17 1BJ, UK \\
$^{16}$Birmingham Institute for Gravitational Wave Astronomy and School of Physics and Astronomy, University of Birmingham, Birmingham B15 2TT, UK
}

\date{Accepted XXX. Received YYY; in original form ZZZ}

\pubyear{2021}

\begin{document}
\label{firstpage}
\pagerange{\pageref{firstpage}--\pageref{lastpage}}
\maketitle

\begin{abstract}
\\
Stripped-envelope supernovae (SE-SNe) show a wide variety of photometric and spectroscopic properties. This is due to the different potential formation channels and the stripping mechanism that allows for a large diversity within the progenitors outer envelop compositions. Here, the photometric and spectroscopic observations of SN\,2020cpg covering $\sim 130$ days from the explosion date are presented. SN\,2020cpg ($z = 0.037$) is a bright SE-SNe with the $B$-band peaking at $M_{B} = -17.75 \pm 0.39$ mag and a maximum pseudo-bolometric luminosity of $L_\mathrm{max} = 6.03 \pm 0.01 \times 10^{42}$ \ergs. Spectroscopically, SN\,2020cpg displays a weak high and low velocity \Ha\ feature during the photospheric phase of its evolution, suggesting that it contained a detached hydrogen envelope prior to explosion. From comparisons with spectral models, the mass of hydrogen within the outer envelope was constrained to be $\sim 0.1$ \msun. From the pseudo-bolometric light curve of SN\,2020cpg a \Nifs\ mass of $M_\mathrm{Ni} \sim 0.27 \pm 0.08$ \msun\ was determined using an Arnett-like model. The ejecta mass and kinetic energy of SN\,2020cpg were determined using an alternative method that compares the light curve of SN\,2020cpg and several modelled SE-SNe, resulting in an ejecta mass of $M_\mathrm{ejc} \sim 5.5 \pm 2.0$ \msun\ and a kinetic energy of $E_\mathrm{K} \sim 9.0 \pm 3.0 \times 10^{51}$ erg. The ejected mass indicates a progenitor mass of $18 - 25$ \msun. The use of the comparative light curve method provides an alternative process to the commonly used Arnett-like model to determine the physical properties of SE-SNe. \\ \\

\end{abstract}

\begin{keywords}
supernovae: general -- supernovae: individual (SN\,2020cpg) \\
\end{keywords}



\section{Introduction}
Core-Collapse Supernovae (CC-SNe) result from the death of stars with a Zero Age Main Sequence (ZAMS) mass of \mzams\ $> 8$ \msun\ \citep{1995ApJ...448..315W, 2009ARA&A..47...63S}. These CC-SNe, separated into multiple categories based on their photometric and spectroscopic properties, are known as the H-rich Type II SNe (SNe\,II) and the H-poor stripped envelope SNe (SE-SNe). SNe\,II undergo little to no stripping of their outer hydrogen envelope prior to explosion and as such display strong hydrogen features throughout their spectral evolution. SE-SNe, however, lack the strong hydrogen features and display a variety of different spectroscopic properties depending on their elemental composition prior to core-collapse. The type of SE-SN can be determined by the presence and strength of both hydrogen and helium features within their spectra. These SNe include the H/He-rich Type IIb SNe (SNe\,IIb), the H-poor/He-rich Type Ib SNe (SNe\,Ib) and the H/He-poor Type Ic SNe (SNe\,Ic). \par

SNe\,Ib(c) lack any prominent hydrogen (and helium) spectral lines \citep{doi:10.1146/annurev.astro.35.1.309}, as their progenitor stars are thought to have been fully stripped of their outer hydrogen and H/He envelopes prior to the core-collapse event. The stripping of the outer envelopes for these progenitor stars is expected to occur over the last stages of the stars life-cycle prior to core-collapse. The process required to strip the outer envelope from these massive stars is still under investigation. The predominant methods include binary interaction where mass is transferred to the companion star via Roche lobe overflow \citep[e.g.][]{1992ApJ...391..246P, 2009MNRAS.396.1699S, 2017MNRAS.470L.102S}, and a single star formation channel where the outer envelope is stripped prior to collapse, during the Wolf-Rayet phase, by either stellar winds \citep[e.g.][]{2012A&A...542A..29G, 10.1093/mnras/stv2283} or via rotational stripping \citep{2013A&A...550L...7G}. Despite the existence of multiple potential formation channels for SE-SNe, the binary star model seems to be favoured in recent years as the dominant source of SE-SNe progenitors. This is because the single star model is unable to produce the number of progenitors required to account for all the SE-SNe observed \citep{10.1111/j.1365-2966.2011.17229.x}.

However, if the degree of stripping is not high enough to fully remove all of the hydrogen from the progenitor, a hydrogen envelope is present during the explosion resulting in a SNe\,IIb \citep[see][for SN\,1993J one of the best followed examples of SN IIb]{1994ApJ...429..300w}. SNe\,IIb are different to the other SE-SNe by the clear hydrogen features within their spectra that can persist for several months before slowly fading as the SN evolves into the nebular phase \citep[see][]{doi:10.1146/annurev.astro.35.1.309, Filippenko:2000yf}. Photometrically, SNe\,IIb are very similar to other SE-SNe displaying a main peak within the first two to three weeks from the explosion. Several SNe\,IIb also display a bright initial peak within a few days of the explosion prior to the main peak seen in all SE-SNe \citep[see][]{2012ApJ...757...31B, 2015ApJ...808L..51P}. The initial luminous peak is thought to result from the shock cooling near the stellar surface \citep{Waxman_2017}, while the second main peak is a result of the radioactive decay of \Nifs\ and \Cofs\ synthesised during the explosion. The dual peak light curve has been seen in several of the well observed SNe\,IIb, such as SN\,1993J \citep{1993ApJ...417L..71W} and SN\,2016gkg \citep{2017ApJ...837L...2A, 2018Natur.554..497B}, and also in some SNe\,Ib, such as SN\,1999ex \citep{2002ApJ...572L..61M} and SN\,2008D \citep{Malesani_2009}. Although this feature is seen in both SNe\,IIb and Ib it is not observed in the majority of SNe, because the progenitor compactness causes the shock to cool more quickly than surveys can observe the shock breakout. However, thanks to the improving cadence of present surveys, the probability of covering and detecting this feature will increase.

Spectroscopically SNe\,IIb differentiate themselves from SNe\,Ib by the presence of the hydrogen features which fade over time as the spectra of SNe\,IIb become more Ib-like, with the helium features becoming dominant. From detailed modelling of H-rich and H-poor SNe the mass of hydrogen within the outer envelope, \mh, required to form a SNe IIb has been found to be within the range of $0.01 - 1.0$ \msun\ \citep{Sravan_2019}. \citet{10.1111/j.1365-2966.2012.20464.x} constructed a detailed set of spectral models to determine the amount of hydrogen and helium that can be hidden within the outer envelope of SNe\,Ib/c respectively. From their synthetic spectra, \citet{10.1111/j.1365-2966.2012.20464.x} concluded that as little as $0.025 - 0.033$ \msun\ is required to form a strong \Ha\ absorption feature, suggesting that some Type Ib's may display \Ha\ features further blending the distinction between IIb and Ib SE-SNe. More recently, \citet{Prentice_2017} showed that the distinction between the He-rich SE-SNe can be further blurred based on the strength of \Ha\ emission within the spectra. \citet{Prentice_2017} created two further SE-SNe subcategories; the Type IIb(I), which display moderate H-rich spectra where the \Ha\ P-Cygni profile is dominated by the absorption component relative to the emission profile, and the Ib(II), whose spectra only show some weak \Ha\ with no obvious Balmar lines more energetic than \Ha. The classification scheme of \citet{Prentice_2017} along with the findings of \citet{10.1111/j.1365-2966.2012.20464.x} demonstrates that SNe\,IIb and Ib are likely more related than previously thought.

\begin{figure}
    \centering
    \includegraphics[width=\columnwidth]{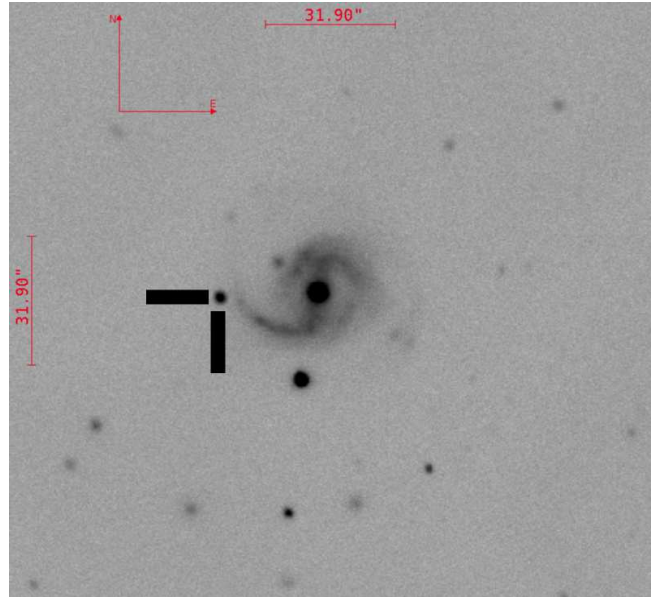}
    \caption{Image of SN\,2020cpg and the host galaxy, obtained by combining LCO observations in $BgVri$ filters on 20/02/2020, stacked and aligned using AstroImageJ \protect\citep{Collins_2017}. The field of view is $2.6 \times 2.4 \enskip \mathrm{arcmin}^{2}$.}
    \label{Location_pic}
\end{figure}

\begin{figure*}
    \centering
    \includegraphics[width=0.9\linewidth]{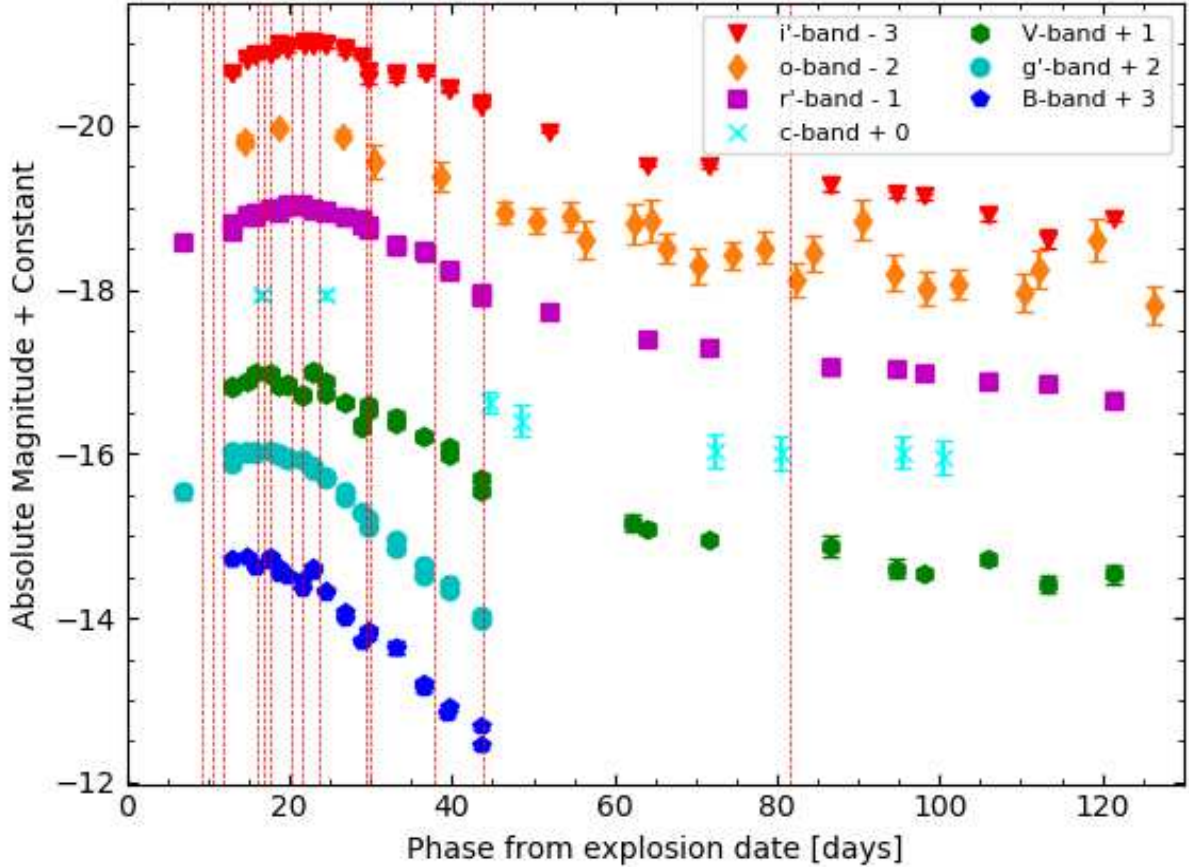}
    \caption{The absolute magnitude photometry of SN\,2020cpg in the $BgVri$-bands along with the ATLAS $c$ and $o$-band covering $\sim 130$ days from the explosion date. The individual band light curves have been corrected for extinction, shifted by a constant magnitude and are shown in rest frame. The red dashed lines denote the epochs at which spectra were taken.}
    \label{BgVri-phot}
\end{figure*}

Here we present the photometric and spectroscopic evolution for SN\,2020cpg, a Type Ib SN with a thin hydrogen layer, during the first $\sim 130$ days. SN\,2020cpg was initially classified with the Supernova Identification code SNID \citep{Blondin_2007} as a Type Ib SN, from the spectrum obtained on 19/02/2020 with the Liverpool Telescope \citep[LT;][]{2004SPIE.5489..679S}. However, follow-up spectral observations suggests that SN\,2020cpg displayed \Ha\ features as seen in Type IIb SNe. In Section \ref{Photometry} we present the $BgVri$-band photometry for SN\,2020cpg from the first 130 days after the explosion obtained through various Las Cumbres Observatory Global Telescope network telescopes \citep[LCO;][]{2013PASP..125.1031B}, as part of the Global Supernova Project \citep[GSP;][]{2017AAS...23031803H}. The spectroscopic observation of SN\,2020cpg are presented in Subsection \ref{Spectroscopy}. In Section \ref{Method} we discuss the construction of the pseudo-bolometric light curve and the Arnett-like model used to obtain the physical parameters. In Section \ref{Results} we present the light curves for the $BgVri$-band photometry and the constructed pseudo-bolometric light curve, along with physical properties obtained by an Arnett-like model. In Section \ref{line evos}, we obtain the line velocity evolution, along with a comparison of SN\,2020cpg spectra with other well followed Type Ib and IIb SNe. In Section \ref{Discussion} we discuss the potential presence of a hydrogen envelope and the spectral modelling done to determine its presence. We also discuss the use of hydrodynamical models to obtain more realistic explosion parameters and compare the results with those produced by the Arnett-like model. Finally, in Section \ref{Sum+Con} we summarise the finding on SN\,2020cpg, giving final estimates for the physical parameters and a value of the progenitors initial mass. 

\section{Observations and Data Reduction}
\label{Obs+DataReduct}
\subsection{Explosion date and Host Galaxy}
\label{Host_galaxy}
SN\,2020cpg was first detected on 15/02/2020 ($\mathrm{MJD} = 58894.54$) by \citet{2020TNSTR.511....1N} on behalf of the Zwicky Transient Facility \citep[ZTF;][]{Bellm_2018}. The last non-detection of SN\,2020cpg, on 06/02/2020 ($\mathrm{MJD} = 58885.52$), predates the ZTF discovery by 9 days. To place a better constraint on the explosion date of SN\,2020cpg we modified the pseudo-bolometric light curve model to include the explosion date as a parameter, see Section \ref{physical_params}. From this fit we obtain an explosion date of 08/02/20, $\mathrm{MJD} = 58887.7 \pm 2.1$ days, which we adopt throughout the rest of the paper.
SN\,2020cpg was associated with the galaxy SDSS J135219.64+133432.9 and was located 1.14” South and 24.07” West from the galaxy centre, just off the outer end of the host galaxy’s western spiral arm, as seen in Figure \ref{Location_pic}. Using the cosmological parameters of H$_{0} = 73.0 \pm 5.0 \enskip \mathrm{km/sec/Mpc}, \enskip \Omega_\mathrm{matter} = 0.27 \text{ and } \Omega_\mathrm{vacuum} = 0.73$ gives a redshift distance of $158.6 \pm 11.1 \enskip \mathrm{Mpc}$, with the distance calculation based on the local velocity field model from \citet{2000ApJ...529..786M}. The host redshift of $z = 0.037$, implies a distance modulus of $m - M = 36.05 \pm 0.15 \enskip \mathrm{mag}$. 

\subsection{Photometry}
\label{Photometry}
The initial $g$ and $r$-band photometry was obtained by ZTF using the ZTF-cam mounted on the Palomar 1.2m Samuel Oschin telescope several days ($t[\mathrm{MJD}] \approx 58894.5$) before continuous follow-up occurred. This photometry was run through the automated ZTF pipeline \citep{Masci_2018} and is presented on Lasair transient broker \citep{2019RNAAS...3...26S} \footnote{https://lasair.roe.ac.uk/object/ZTF20aanvmdt/}. After the discovery, the $BgVri$-bands were followed by the Las Cumbres Observatory Global Telescope network \citep[LCO;][]{2013PASP..125.1031B} and reduced using the BANZAI pipeline \citep{10.1117/12.2314340}. Full $BgVri$-band photometry was obtained until 23/03/2020 from which point only $Vri$-band photometry could be obtained. Observation was obtained from a combination of 1\,m telescopes from the Siding Spring Observatory (code: COJ), the South African Astronomical Observatory (code: CPT), the McDonald Observatory (code: ELP) and the Cerro Tololo Interamerican Observatory (code: LSC). Both $c$ and $o$-band photometry were also obtained by the Asteroid Terrestrial-impact Last Alert System \citep[ATLAS;][]{Smith_2020} and reduced through the standard ATLAS pipeline \citep{Tonry_2018}. The $BgVri + co$-band absolute light curve from the follow-up campaigns are shown in Figure \ref{BgVri-phot}. The photometry has been corrected for reddening using a Milky Way (MW) extinction of $E(B - V)_{\mathrm{MW}} = 0.025 \pm 0.001$ mag, obtained using the Galactic dust map calibration of \citet{2011ApJ...737..103S} and extinction factor $R_{V} = 3.1$. The host galaxy extinction was taken to be negligible relative to MW extinction, as there was no noticeable \NaI\ D $\lambda \lambda\,5890, 5896$ lines at the SN rest frame, \citep[e.g.][]{10.1111/j.1365-2966.2012.21796.x}. Also it should be noted that, as seen in Figure \ref{Location_pic}, SN\,2020cpg was located far from the galactic centre where the effect of dust is likely reduced. All uncorrected LCO photometry and ATLAS photometry are given in Table \ref{All_photometry} and Table \ref{ATLS_photometry} respectively. 
 
\subsection{Spectroscopy}
\label{Spectroscopy}

\begin{figure*}
    \centering
    \includegraphics[width=0.9\linewidth, height=20cm]{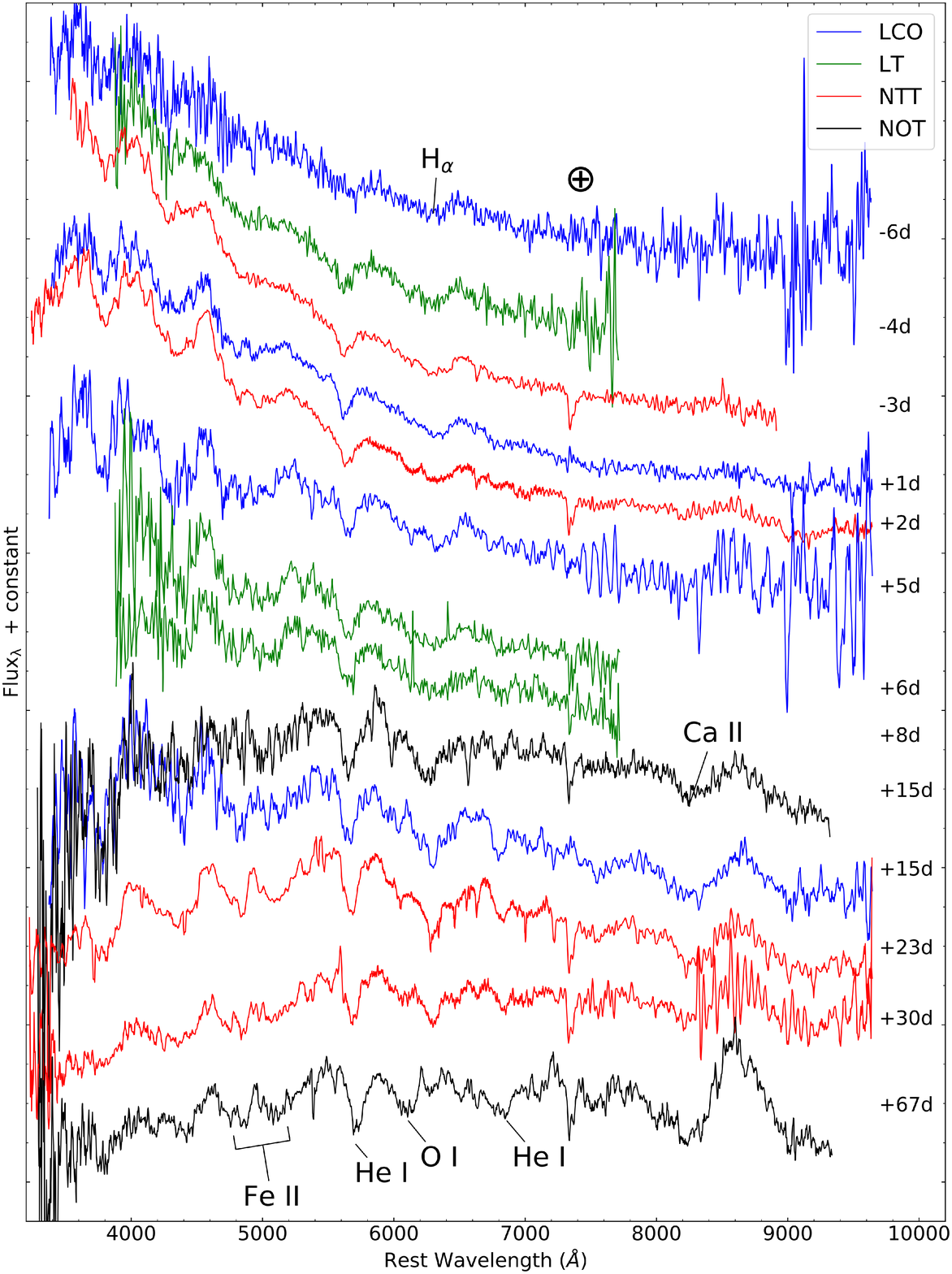}
    \caption{Spectroscopic evolution of SN\,2020cpg with details of the observations given in Table \ref{Spectra epochs}. The epochs on the right side are relative to $B_{max}$ in rest frame. The \Ha, \HeI, \FeII, \CaII\ and \OI\ features have been noted along with the main telluric feature at 7600, $\oplus$. The spectra have been binned to reduce the noise.}
    \label{All_spec}
\end{figure*}

Spectra from multiple telescopes were obtained over an 80 day period post explosion and reduced through standard means available within each observatory pipeline. The classification spectrum of SN\,2020cpg \citep{2020TNSCR.571....1P} was obtained with the LT, on 19/02/2020 using the Spectrograph for the Rapid Acquisition of Transients \citep[SPRAT;][]{10.1117/12.2055117} and was reduced by the LT automatic pipeline \footnote{http://telescope.livjm.ac.uk/TelInst/Inst/SPRAT/} \citep[see][for details on the pipeline]{2012AN....333..101B}. Several later spectra were also obtained using the LT. Additional spectra for SN\,2020cpg were obtained by the advanced Public ESO Spectroscopic Survey for Transient Objects (ePESSTO+) \footnote{www.pessto.org} \citep[][]{2015A&A...579A..40S} using the ESO Faint Object Spectrograph and Camera mounted on the New Technology Telescope (NTT) \citep[EFOSC2;][]{1984Msngr..38....9B}. ePESSTO+ data were reduced as described in \citet{2015A&A...579A..40S}. The Alhambra Faint Object Spectrograph and Camera (ALFOSC) mounted on the Nordic Optical Telescope \citep[NOT;][]{Djupvik_2010} provided several spectra of SN\,2020cpg, which were reduced by the Foscgui pipeline \footnote{http://graspa.oapd.inaf.it/foscgui.html}. Multiple spectra were taken by the LCO 2\,m Faulkes Telescope South (FTS) at COJ and Faulkes Telescope North (FTN) at the Haleakala Observatory (code: OGG). We attempted to obtain further spectra after two and a half months post explosion, however SN\,2020cpg was too dim at this point for the available telescopes to obtain good quality spectra. All spectra have been binned to improve the S/N ratio, and de-reddened, assuming a standard $R_{V} = 3.1$ and the $E(B - V)$ given in Section \ref{Host_galaxy}. All spectra can be seen in Figure \ref{All_spec}. The details on the phase from $B$-band max, observatory and instrument alongside the observed range are given in Table \ref{Spectra epochs}. 

\begin{table}
    \caption{Details of the spectroscopic observations of SN\,2020cpg. Phase from both the predicted explosion date (Phase$_\mathrm{exp}$) and the date of $B_\mathrm{max}$ (Phase$_{B_\mathrm{max}}$) are given in rest-frame.}
    \centering
    \begin{tabular}[width=\textwidth]{ccccc}
    Date & Phase$_\mathrm{exp}$ & Phase$_{B_\mathrm{max}}$ & Telescope + & Range \\
    & [Days] & [Days] & Instrument &[$\si\angstrom$] \\
    \hline
    17/02 & +9 & -6 & FTS en12 & 3500 - 10000 \\
    19/02 & +11 & -4 & LT SPRAT & 4000 - 8000 \\
    20/02 & +12 & -3 & NTT EFOSC2 & 3685 - 9315 \\
    24/02 & +16 & +1 & FTN FLOYDS & 3500 - 9000 \\
    25/02 & +17 & +2 & NTT EFOSC2 & 3380 - 10320 \\
    28/02 & +20 & +5 & FTS en12 & 3500 - 10000 \\
    29/02 & +21 & +6 & LT SPRAT & 4000 - 8000 \\
    02/03 & +23 & +8 & LT SPRAT & 4000 - 8000 \\
    09/03 & +30 & +15 & NOT ALFOSC & 3200 - 9600 \\
    09/03 & +30 & +15 & FTN FLOYDS & 3500 - 10000 \\
    17/03 & +38 & +23 & NTT EFOSC2 & 3380 - 10320 \\
    23/03 & +44 & +30 & NTT EFOSC2 & 3380 - 10320 \\
    30/04 & +82 & +67 & NOT ALFOSC & 3200 - 9600 \\
    \hline
    \end{tabular}
    \label{Spectra epochs}
\end{table}

\section{Method}
\label{Method}

\subsection{Pseudo-bolometric Light curve}
\label{Bol_model}

\begin{figure}
    \centering
    \includegraphics[width=1\columnwidth]{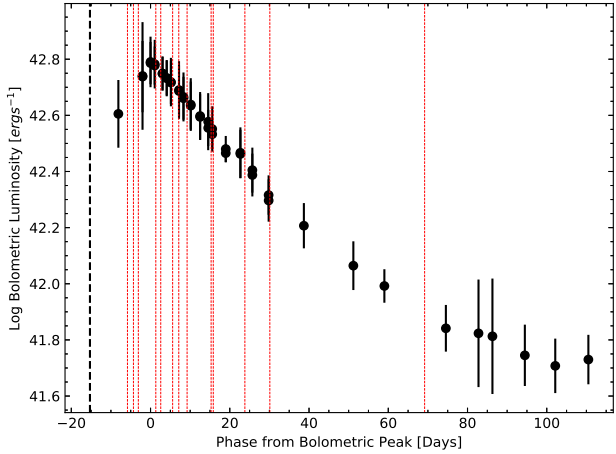}
    \caption{The pseudo-bolometric light curve of SN\,2020cpg constructed from the $BgVri$ photometry. Luminosity is shown relative to days from the peak of the pseudo-bolometric light curve in rest-frame and follows approximately 120 days from explosion. The red dashed lines indicate the epochs where spectra were taken and the black dashed line is the yielded explosion date.}
    \label{pseudo-bol}
\end{figure}

From the $BgVri$-band photometry obtained for SN\,2020cpg we constructed a pseudo-bolometric light curve, shown in Figure \ref{pseudo-bol}, using the pseudo-bolometric light curve code of \citet{Nicholl_2018}. As we lack any UV or NIR data, we approximate the missing luminosity in these bands by extrapolating the blackbody spectral energy distributions that were fit to the $BgVri$-bands into the UV and NIR regions. The UV and NIR contributions to the pseudo-bolometric light curve are relatively small at peak time, contributing $\sim 10 - 20 \%$ and $\sim 15 - 25 \%$ respectively, compared to the optical contribution, which accounts for $\sim 50 - 60\%$ of total flux near bolometric peak \citep{10.1093/mnras/stt2187}. We conclude that our extrapolation to UV and NIR bands does not introduce a significant error to the bolometric light curve. 

Along with the pesudo-bolometric light curve of SN\,2020cpg, we construct pesudo-bolometric light curves for SN\,1993J \citep{1994AJ....107.1022R, 1995A&AS..110..513B, 1996AJ....112..732R}, SN\,2003bg \citep{Hamuy_2009}, SN\,2008ax \citep{2008MNRAS.389..955P, 2009PZ.....29....2T}, SN\,2009jf \citep{2011MNRAS.413.2583S, 2014ApJS..213...19B}, SN\,2011dh \citep{tsvetkov2012photometric, 2013MNRAS.433....2S, 2014Ap&SS.354...89B}, iPTF13bvn \citep{2014Ap&SS.354...89B, 2016A&A...593A..68F, 2016ApJ...825L..22F}, SN\,2013ge \citep{Drout_2016}, 2015ap \citep{2019MNRAS.485.1559P} and SN\,2016gkg \citep{2014Ap&SS.354...89B, 2017ApJ...837L...2A, 2018Natur.554..497B}. The comparison between these SE-SNe is shown in Figure \ref{all_bol+fit}. These SE-SNe were chosen as they all have comprehensive coverage over the first $\sim 100$ days post explosion, they all have well defined explosion dates and photospheric velocities both of which are required for the Arnett-like model used to obtain physical parameters.
For these SE-SNe, we excluded any UV and NIR data available when constructing the pseudo-bolometric light curve ensuring the effects of the UV and NIR extrapolation did not greatly influence the comparison between the SE-SNe. Where SNe lacked Sloan Digital Sky Survey (SDSS) filters we used the corresponding Johnson-Cousins (J-C) filters to cover a similar wavelength range allowing for a more accurate comparison between the pseudo-bolometric light curves.

\subsection{Physical parameters}
\label{physical_params}

The bolometric luminosity of a SN is intrinsically linked to several physical parameters, those being the mass of nickel synthesised during the explosion, the amount of material ejected from the outer layers of the progenitor and the kinetic energy of the ejected mass. This relation was first formulated for Type Ia SNe by \citet{1982ApJ...253..785A} who assumed that all the energy that powers the bolometric light curve originated from the decay of $\prescript{56}{}{Ni} \rightarrow \prescript{56}{}{Co}$ and the decay of $\prescript{56}{}{Co} \rightarrow \prescript{56}{}{Fe}$. While the model was initially formulated for SNe that do not undergo a hydrogen recombination phase, such as those seen in SE-SNe\,Ib/c and SNe\,IIb, it has been used regularly for multiple types of SNe. This is done by ignoring the recombination phase and restricting the fitting to the rise and fall of the peak of the bolometric light curve that is powered by radioactive activity, as done in \citet[][]{Lyman_2016}. The Arnett-like model also assumes that all \Nifs\ is located in a point at the centre of the ejecta, that the optical depth of the ejecta is constant throughout the evolution of the light curve, the initial radius prior to explosion is very small and that the diffusion approximation used for the model is that of photons. While these assumptions are acceptable, the approximation of constant opacity has a severe effect on diffusion timescale which is dependent on the estimated ejecta mass and kinetic energy of the SN. The effect of neglecting the time-dependent diffusion on the \Nifs\ mass was discussed by \citet{Khatami_2019}, who concluded that this results in an over estimation of the \Nifs\ mass. Through alternative modelling methods it was seen that the \Nifs\ mass was overestimated by the Arnett-like model by $\sim 30 - 40\%$ \citep[see,][]{10.1093/mnras/stw418, 2020arXiv200906868W}. \par
We initially used the Arnett-like model to determine the physical parameters of SN\,2020cpg and compare the results to several other SE-SNe. In the Arnett-like model the kinetic energy and ejecta mass have a strong dependence on the diffusion timescale, $\tau_\mathrm{m}$, of the bolometric light curve, which is given as;
\begin{ceqn}
\begin{align}
    \centering
    \label{tau_m}
    \tau_\mathrm{m} = \left(\dfrac{\kappa_\mathrm{opt}}{\beta c}\right)^{\dfrac{1}{2}} \left(\dfrac{6 M_\mathrm{ejc}^3}{5 E_\mathrm{k}}\right)^{\dfrac{1}{4}}.
\end{align}
\end{ceqn}
Where $M_\mathrm{ejc}$ is the mass of ejected material and $E_\mathrm{k}$ is the kinetic energy of the supernovae. Also $c$ is the speed of light, $\beta$ is the constant of integration derived by \citet{1982ApJ...253..785A} that takes the value of $\beta \approx 13.8$ and $\kappa_\mathrm{opt}$ is the optical opacity of the material ejected by the SN. For the Arnett-like model a constant value of $\kappa_\mathrm{opt} = 0.06 \pm 0.01 \mathrm{cm^{2} g^{-1}}$ was used. The degeneracy between the ejecta mass and kinetic energy was broken using the photospheric velocity the event obtained from the velocity of the \FeII\ $5169 \si{\angstrom}$ line measured at maximum bolometric luminosity. This is the epoch when the outer ejecta has the largest contribution to the luminosity under the assumption of homogeneous density. The model was also adjusted to include the SN explosion date to allow for an improved fit and to place a constraint on the rise time of the SNe. For SNe with well observed pre-maximum and well defined explosion dates we use the dates provided. The explosion date of SN\,2020cpg was obtained by constraining the fitting to limit the potential explosion date to after the date of last non-detection and prior to the initial observation. \par

Due to the known problems with the Arnett-like model, in Section \ref{ratio_properties} we discuss an alternative method for determining the ejecta mass and kinetic energy of SN\,2020cpg by comparing the light curve properties and physical properties determined by hydrodynamical modelling of other SE-SNe, as done for SN\,2010ah in \citet[][here after PM13]{Mazzali_2013}. This method re-scales the physical parameters of other SE-SNe using equation \ref{tau_m} under the assumption that the optical opacity of the two SNe are equivalent. This is physically a more robust assumption than a fixed opacity for all SE-SNe as adopted by the Arnett-like model. A comparison between the results obtained from the Arnett-like model and the PM13 model is presented later in Section \ref{ratio_properties}. \par

\section{Results}
\label{Results}
\subsection{Multi-colour light curves}

\begin{table}
    \caption{Epoch of light curve maximum, rise time in rest-frame and peak absolute magnitude for the $BgVri$ photometry bands for SN\,2020cpg}
    \centering
    \begin{tabular}{c@{\hspace{0.5\tabcolsep}}c@{\hspace{0.5\tabcolsep}}c@{\hspace{0.75\tabcolsep}}c@{\hspace{0.75\tabcolsep}}c}
        Band & $MJD_\mathrm{max}$ & Rise time (days) & $m_\mathrm{max}$ & $M_\mathrm{max}$ \\
        \hline
        $B$ & 58902.1 & $14.7 \pm 2.5$ & $18.38 \pm 0.02$ & $-17.75 \pm 0.39$ \\
        $g'$ & 58903.1 & $16.0 \pm 2.1$ & $18.05 \pm 0.02$ & $-18.04 \pm 0.40$ \\
        $V$ & 58904.7 & $17.0 \pm 2.1$ & $18.16 \pm 0.02$ & $-17.91 \pm 0.38$ \\
        $c$ & 58906.0 & $18.8 \pm 2.3$ & $18.10 \pm 0.05$ & $-17.97 \pm 0.38$ \\
        $r'$ & 58906.2 & $18.6 \pm 2.1$ & $18.06 \pm 0.02$ & $-18.00 \pm 0.38$ \\
        $o$ & 58908.3 & $21.1 \pm 2.4$ & $18.05 \pm 0.05$ & $-18.01 \pm 0.39$ \\
        $i'$ & 58909.2 & $22.0 \pm 2.1$ & $18.05 \pm 0.02$ & $-18.00 \pm 0.38$ \\
        \hline
    \end{tabular}
    \label{peak_mag}
\end{table}

The early time rise of both the $B \text{ and } V$-bands were missed in the follow-up campaign, however the peaks in both bands were observed, shown in Figure \ref{BgVri-phot}. The bluer bands peaked several days before the red bands, $t_\mathrm{blue}^\mathrm{rise} \approx 15 \text{ days post explosion and }t_\mathrm{red}^\mathrm{rise} \approx 19 \text{ days}$. Both the $B$ and $g$-bands were followed for $\sim 30$ days by LCO before the photometry bands dropped below the brightness threshold required for follow-up. The brightness for the $B$ and $g$-bands fell by $\sim 2$ magnitudes in the 30 days from the photometric peak as a result of the SN rapidly cooling. The remaining bands fell at a slower rate, dropping by roughly 1 magnitude in the same time period, before their decline slowed down as the light curve transitioned to the exponential tail produced by the radioactive decay \Cofs\ synthesised in the explosion.
The ATLAS $c$-band was followed for approximately 100 days from the expected explosion date with the $o$-band being followed for a further 30 days. The peaks in both bands were not well observed, especially in the $c$-band. As with the other bands the redder $o$-band declines at a slower rate just after maximum light when compared to the $c$-band. The ATLAS bands have a greater error associated with them compared to the $BgVri$-bands, and as the ATLAS bands cover a similar wavelength range as the $BgVri$ they were not used when constructing the pseudo-bolometric light curve. \par

\begin{figure*}
    \centering
    \includegraphics[width=\linewidth]{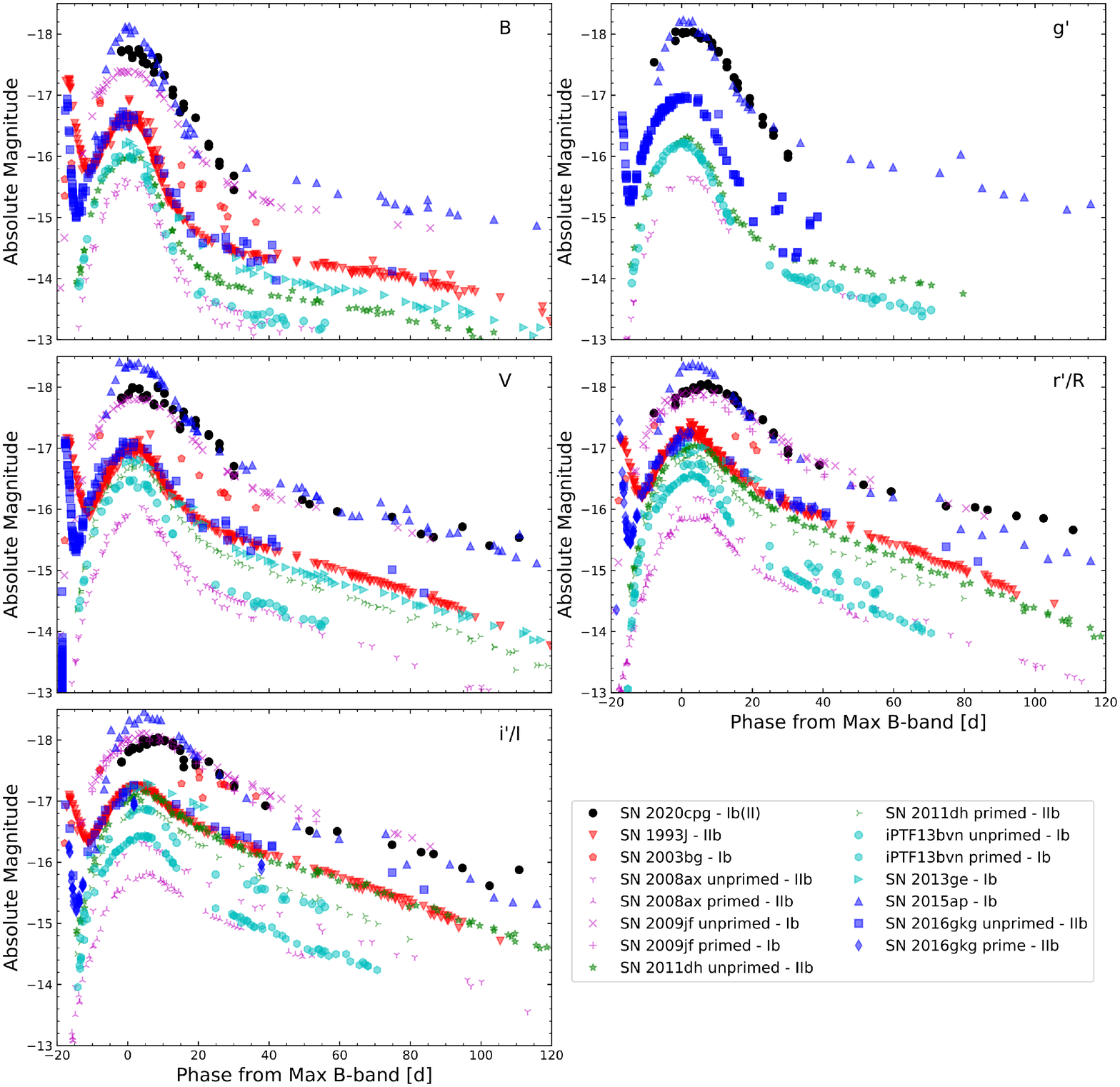}
    \caption{Comparison of the absolute magnitude light curves of several SNe\,Ib and IIb with SN\,2020cpg. All photometry is relative to $B_\mathrm{max}$ light which was either taken from the literature or by fitting a Gaussian to the $B$-band peak. The light curves have been corrected for time dilation as well as corrected for both Milky Way and host galaxy reddening when possible. Primed bands are SDSS photometry bands and unprimed are the Johnson-Cousins photometry bands. Error on absolute magnitudes not included.}
    \label{abs_B_date}
\end{figure*}

The light curves for He-rich CC-SNe display a variation within the evolution of their light curves due to the range of progenitor properties. As such the $BgVri$-band photometry for SN\,2020cpg was compared with those of SN\,1993J, SN\,2003bg, SN\,2009jf, SN\,2011dh, iPTF13bvn, SN\,2013ge, SN\,2015ap and SN\,2016gkg. 
The absolute magnitude photometry for these SNe relative to SN\,2020cpg is shown in Figure \ref{abs_B_date}, with the details on each SN given in Table \ref{SN_phot_details}. SN\,2020cpg is brighter than the majority of the other SNe that we compare to, with only SN\,2009jf and SN\,2015ap being of similar brightness. The $B \text{ and } g$-bands evolve in a similar way to that of SN\,2015ap while the other bands evolve more similar to SN\,2009jf. Due to the lack of pre-maximum light observations it is not possible to determine if SN\,2020cpg had a shock breakout cooling peak similar to that seen in several other SE-SNe, such as SN\,1993J and SN\,2016gkg.

 \subsection{pseudo-bolometric Light Curves}
\label{Bol_section}

\begin{figure}
    \centering
    \includegraphics[width=1\columnwidth]{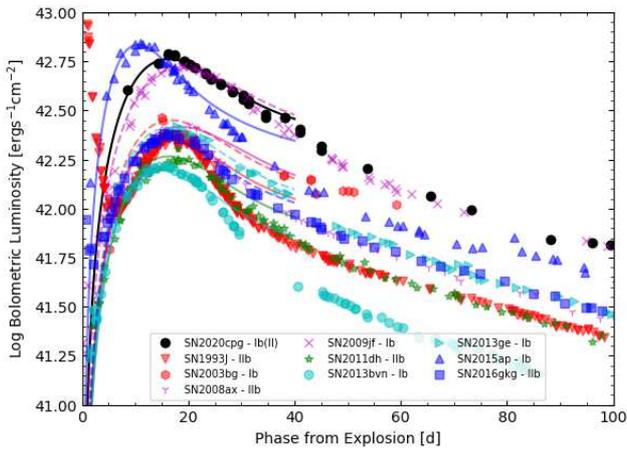}
    \caption{Pseudo-bolometric light curves of SN\,1993J, SN\,2003bg, SN\,2008ax, SN\,2009jf, SN\,2011dh, iPTF13bvn, SN\,2013ge, SN\,2015ap, SN\,2016gkg and SN\,2020cpg covering a period of 100 days from their estimated explosion date. The Arnett-like model fit to the pseudo-bolometric light curves, detailed in Section \ref{physical_params}, are shown as lines and were fitted out to $\sim 40$ days before they started to strongly diverge from the pseudo-bolometric light curves. The velocities used to break the degeneracy for each SN, along with the predicted physical parameters, are given in Table \ref{Arnett_fit_params}.}
    \label{all_bol+fit}
\end{figure}

The pseudo-bolometric rise time for SN\,2020cpg is $t_\mathrm{rise}^\mathrm{bol} \approx 16.0 \pm 2.5$ days. Once peak luminosity had been reached the light curve rapidly declines for the next $\approx 34$ days before settling on the exponential tail. Due to lack of much pre-peak photometry the rise of the pseudo-bolometric light curve is not as well constrained as the post-peak light curve. SN\,2020cpg reaches a peak luminosity of log($L_{\mathrm{max}}) = 42.78 \pm 0.08$\,[\ergs], which is higher than the average luminosity of Type IIb + Ib(II), which has a value of log($L_{\mathrm{max}}) = 42.2 ^{+0.4}_{-0.1}$\, [\ergs], and the average maximum luminosity of Type IIb + IIb(I), log($L_{\mathrm{max}}) = 42.09 \pm 0.17$\, [\ergs], 
as given in \citet{2019MNRAS.485.1559P}, showing that SN\,2020cpg lies at the brighter end of the SE-SNe regime. \par

We fit the pseudo-bolometric light curve of SN\,2020cpg with the Arnett-like model using a photospheric velocity of \vph\ $ \approx 12500 \pm 1500$ \kms\ to break the degeneracy between the kinetic energy and ejecta mass. The value of \vph\ was obtained from the average \FeII\ line velocities at peak light. The average value of the \FeII\ triplet was used instead of the commonly employed \FeII\ \lam\ $5169$ line due to the low signal to noise ratio within the \FeII\ region of the spectrum taken around peak luminosity. From the Arnett-like model fit to SN\,2020cpg’s pseudo-bolometric light curve we derive a nickel mass of \mni\ $= 0.27 \pm 0.08$ \msun. The ejecta mass and kinetic energy given by the fit had a value of \mej\ $= 3.4 \pm 1.0$ \msun\ and \ek\ $=2.9 \pm 0.9 \times 10^{51}$ \erg respectively. This process was then repeated for the bolometric light curves of the other SE-SNe shown in Figure \ref{all_bol+fit} and the derived physical parameters are given in Table \ref{Arnett_fit_params}. As expected the Arnett-like model deviates from the pseudo-bolometric light curves at later times ($t \gtrsim 40$ days) when the SNe start to transition into the nebular phase. Relative to the other SNe, SN\,2020cpg has a high nickel mass similar to both SN\,2009jf and SN\,2015ap, shown in Table \ref{Arnett_fit_params}. The similar \mni\ between SN\,2020cpg and both SN\,2009jf and SN\,2015ap is expected from their comparable peak luminosities. The ejecta mass and kinetic energy of SN\,2020cpg is also higher than the majority of the SE-SNe we have looked at, suggesting that the progenitor of SN\,2020cpg was a high mass star prior to the stripping of the outer envelope. However, due to the problems associated with the Arnett-like approach, we discuss an alternative approach to obtain the values for \mej\ and \ek\ in Section \ref{ratio_properties}, we then use the values for \mej\ and \ek\ derived using the PM13 method to estimate the progenitor mass.

\begin{table*}
    \caption{Details for several historical Type Ib and IIb SNe which have been compared to SN\,2020cpg. Sources- 1: \protect\citet{1994AJ....107.1022R}, 2: \protect\citet{1995A&AS..110..513B}, 3: \protect\citet{1996AJ....112..732R}, 4: \protect\citet{2008MNRAS.389..955P}, 5: \protect\citet{Hamuy_2009}, 6: \protect\citet{2009PZ.....29....2T}, 7: \protect\citet{2011MNRAS.413.2583S}, 8: \protect\citet{tsvetkov2012photometric}, 9: \protect\citet{2013MNRAS.433....2S}, 10: \protect\citet{2014ApJS..213...19B}, 11: \protect\citet{2014Ap&SS.354...89B}, 12: \protect\citet{2016A&A...593A..68F}, 13: \protect\citet{2016ApJ...825L..22F}, 14: \protect\citet{Drout_2016}, 15: \protect\citet{2017ApJ...837L...2A}, 16: \protect\citet{2018Natur.554..497B} and 17: \protect\citet{2019MNRAS.485.1559P}.}
    \centering
    \begin{tabular}{cccccccc}
        SN & Explosion date & $B_\mathrm{max}$ date & Redshift & Distance & $E(B - V)_\mathrm{MW}$ & $E(B - V)_\mathrm{Host}$ & Source \\
        & [MJD] & [MJD] & & [Mpc] & [mag] & [mag] &\\
        \hline
        1993J & 49072.0 & 49093.48 & -0.000113 & 2.9 & 0.069 & 0.11 & 1,2,3\\
        2003bg & 52695.0 & 52718.35 & 0.00456 & 20.25 & 0.018 & - & 5\\
        2008ax & 54528.8 & 54546.86 & 0.001931 & 5.1 & 0.0188 & 0.28 & 4,6\\
        2009jf & 55101.33 & 55120.91 & 0.0079 & 31 & 0.097 & 0.03 & 7,10\\
        2011dh & 55712.5 & 55730.82 & 0.001638 & 7.25 & 0.0309 & 0.05 & 8,9,11\\
        iPTF13bvn & 56458.17 & 56474.95 & 0.00449 & 19.94 & 0.0436 & 0.17 & 11,12,13\\
        2013ge & 56602.5 & 56618.93 & 0.004356 & 19.342 & 0.0198 & 0.047 & 14\\
        2015ap & 57270.0 & 57283.0 & 0.01138 & 50.082 & 0.037 & - & 17\\
        2016gkg & 57651.15 & 57669.67 & 0.0049 & 21.8 & 0.0166 & 0.09 & 11,15,16\\
        2020cpg & 58887.6 & 58902.07 & 0.037 & 158.6 & 0.0246 & - & - \\
        \hline
    \end{tabular}
    \label{SN_phot_details}
\end{table*}
\begin{table}
    \caption{Physical properties of several SE-SNe derived from the fitting of the Arnett-like model described in Section \ref{Method} and shown in Figure \ref{all_bol+fit}. The photospheric velocity used for each SN was taken from their discovery paper.}
    \centering
    \begin{adjustbox}{width=1\columnwidth}
    \begin{tabular}{ccccc}
        SN & \vph & \mni & \mej & \ek \\
         & [\kms] & [\msun] & [\msun] & $[10^{51} \mathrm{erg}]$ \\
        \hline
        1993J & $8000 \pm 1000$ & $0.11 \pm 0.03$ & $ 2.0 \pm 0.8$ & $0.8 \pm 0.3$ \\
        2003bg & $10000 \pm 500$ & $0.13 \pm 0.04$ & $2.8 \pm 0.9$ & $1.7 \pm 0.5$ \\
        2008ax & $7500 \pm 500$ & $0.14 \pm 0.04$ & $2.6 \pm 1.0$ & $0.9 \pm 0.3$ \\
        2009jf & $11000 \pm 500$ & $0.27 \pm 0.09$ & $4.0 \pm 1.4$ & $2.9 \pm 1.0$ \\
        2011dh & $7000 \pm 1000$ & $0.09 \pm 0.02$ & $1.8 \pm 0.5$ & $0.5 \pm 0.1$ \\
        iPTF13bvn & $8000 \pm 1000$ & $0.07 \pm 0.02$ & $1.7 \pm 0.4$ & $0.6 \pm 0.2$ \\
        2013ge & $10500 \pm 500$ & $0.12 \pm 0.03$ & $2.7 \pm 0.8$ & $1.8 \pm 0.5$ \\
        2015ap & $16000 \pm 1000$ & $0.22 \pm 0.05$ & $1.4 \pm 0.4$ & $2.2 \pm 0.6$ \\
        2016gkg & $8000 \pm 1000$ & $0.10 \pm 0.02$ & $1.9 \pm 0.4$ & $0.7 \pm 0.2$ \\
        2020cpg & $12500 \pm 1200$ & $0.27 \pm 0.08$ & $3.4 \pm 1.0$ & $2.9 \pm 0.9$ \\
        \hline
    \end{tabular}
    \end{adjustbox}
    \label{Arnett_fit_params}
\end{table}

\subsection{Spectral Evolution and Comparison}
\label{line evos}

At early times, the spectra of SN\,2020cpg (Figure \ref{All_spec}) shows a large blue excess. The spectra rapidly cool until around +15 days from $B_\mathrm{max}$. Prominent \HeI\ lines are present throughout the spectral evolution with the \HeI\, $5876 \, \si{\angstrom}$ line being the most prominent and the $6678 \, \si{\angstrom}$ line becoming stronger at around +23 days. Around +1 days post $B_\mathrm{max}$ the spectrum develops an absorption feature located in the \Ha\ region which persists for $\sim 30$ days. At earlier times during the spectral evolution, the \Ha\ feature is split into a high velocity and low velocity component which merge into a single \Ha\ feature at later times. The presence of the \Ha\ line provides strong evidence that SN\,2020cpg is not a standard Type Ib SN and may be an intermediate SN between the H-rich and H-poor SE-SNe. 
While the feature around $6300 \si{\angstrom}$ may be interpreted as the presence of silicon, this is not likely, because it would imply that absorption from other silicon transitions, around $4100$ and $5900 \si{\angstrom}$ should be detected in this and later spectra which is not observed. Moreover, when identified as silicon, the line shift would indicate a velocity of $3000$ \kms\ which is far too slow for this epoch.
These pieces of evidence along side the lack of silicon in the spectra of other well observed Type Ib/c SNe gives strong evidence that the feature is the result of the presence of hydrogen within the outer envelope. Later, the spectral evolution shows the development of \FeII\ $\lambda\lambda 4924, 5018, 5169$ lines, although it should be noted that the \FeII\ lines are located close to \HeI\ lines making the separation of these lines difficult, especially given the high noise in this region of the spectra. \par

\begin{figure*}
    \centering
    \includegraphics[width=0.9\linewidth]{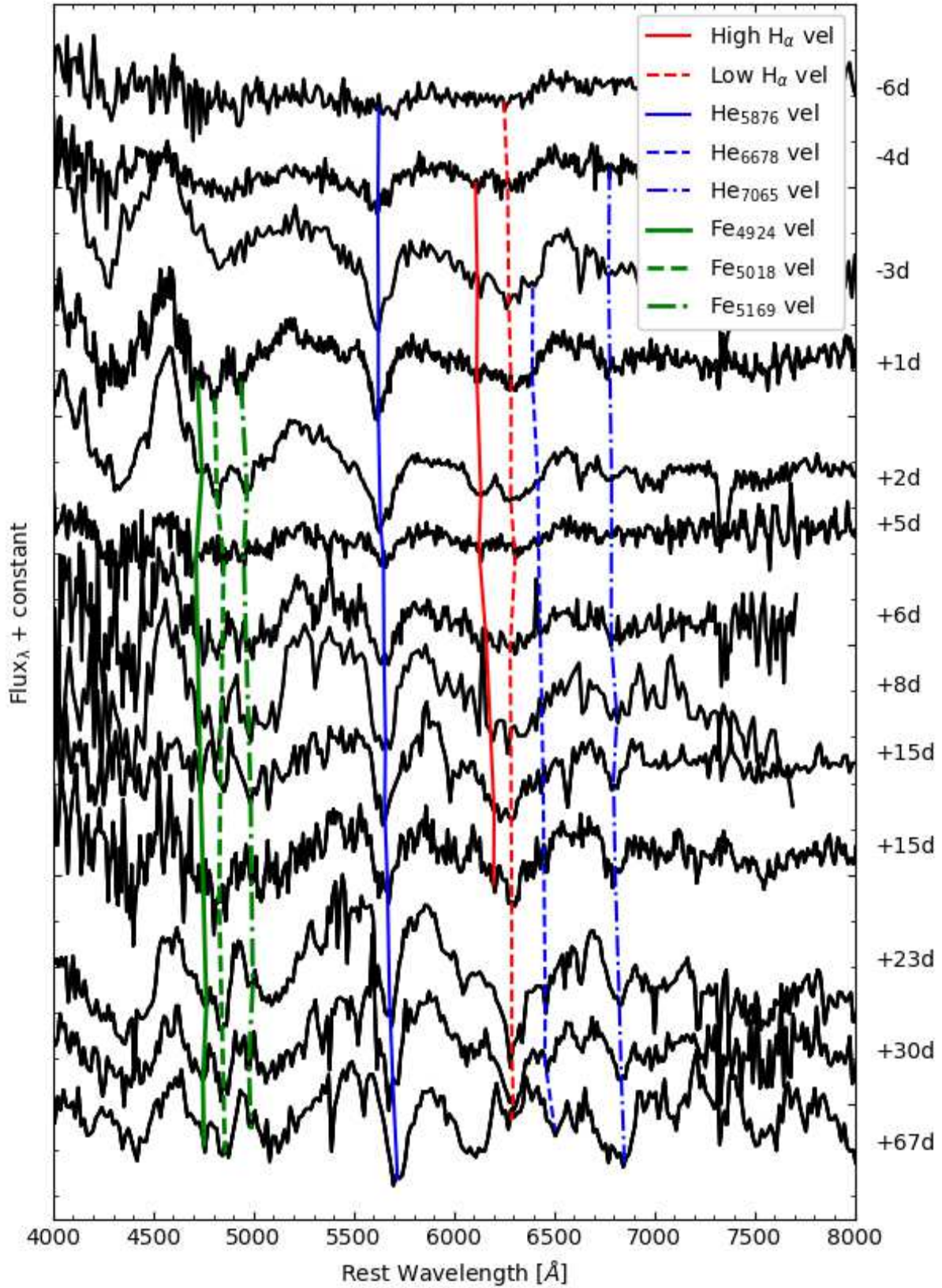}
    \caption{Evolution of SN\,2020cpg spectra. The spectra have been plotted between $4000 - 8000 \, \si{\angstrom}$ to highlight the regions where prominent \Ha, \HeI\ and \FeII\ features are visible. The different elements are shown by the different lines, with \Ha\ = red, \HeI\ = blue and \FeII\ = green, and different element lines given by different styles. Lines are only shown when line features are clearly visible within the spectra.}
    \label{spec+lines}
\end{figure*}

\begin{figure}
    \centering
    \includegraphics[width=\linewidth]{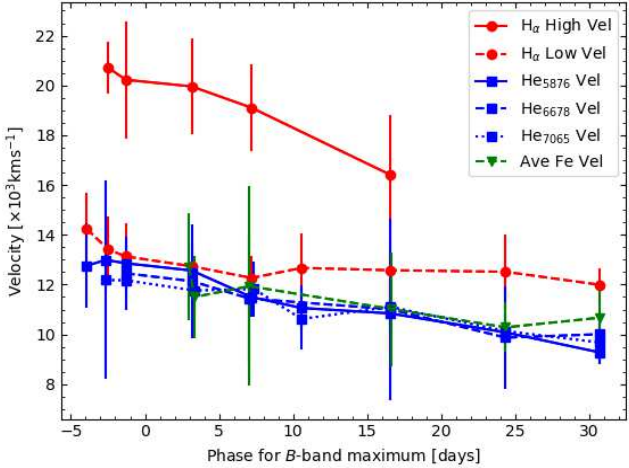}
    \caption{Line velocity evolution of \Ha, \HeI\ and \FeII\ from Gaussian fits to the spectra. The \Ha\ (red) is split into a high velocity and a low velocity component, shown as the solid line and dashed line respectively. The evolution of a selection of individual \HeI\ spectral lines (blue) are shown as separate curves. Due to the uncertainty in the \FeII\ lines (green), as a result of the noise in that region of the spectrum, the average line velocity is displayed. The plot is cut off at $\sim 30$ days due to the emergence of other lines around the \Ha\ region.}
    \label{Line_vel_evo}
\end{figure}

The evolution of the line velocities for \Ha, \HeI\ $\lambda\lambda 5876, 6678, 7065$ and \FeII\ $\lambda\lambda 4924, 5018, 5169$ were determined by the fitting of a Gaussian to each feature to locate the minima. The line evolution of each elemental feature is shown in Figure \ref{spec+lines}. The line velocities derived from the Gaussian fits are given in Figure \ref{Line_vel_evo}. The main source of error for these elemental line velocities comes from the low S/N of the spectra, especially on the fringes where the \FeII\, line is located, which makes the fitting of the Gaussian more difficult. This results in an error derived from the Gaussian fitting of approximately 15\%, with a negligible error associated with the redshift. 
For the \Ha\ feature, we separate the minimum into two distinct high and low velocity components. The high velocity feature is visible from the second spectrum, $-4$ days, until approximately $+15$ days post $B_\mathrm{max}$, as shown by the solid red line in Figure \ref{spec+lines}. At this point the high velocity and low velocity components blend together in the later spectra to form a single \Ha\ feature. There is a clear separation between the high and low velocity \Ha\ components, with the low velocity remaining relatively constant in velocity with a decline of $\sim 2000$ \kms\ from $\sim 14500$ \kms\ to $\sim 12500$ \kms\ while the high velocity component drops by $\sim 5000$ \kms from $\sim 21000$ \kms\ to $\sim 16500$ \kms\ before the lines seem to merge into one constant \Ha\ feature. The \HeI\ \lam $5678$ feature remains strong throughout the spectral evolution while the \HeI\ \lam $6678$ feature, while not always visible due to high noise, follows the velocity evolution of \HeI\ \lam $5876$.
The signal to noise ratio in the \FeII\ region made finding the velocity evolution harder than for the other lines. The velocity of the \HeI\ and \FeII\ lines all follow a similar trend declining from $\sim 13000$ \kms to $\sim 10000$ \kms. From the velocity evolution we determine that the photospheric velocity of SN\,2020cpg at peak luminosity has an average value of $12500 \pm 1200$ \kms, taken from the velocity of the \HeI\ and average \FeII\ features. \par

\begin{figure*}
    \centering
    \includegraphics[width=\linewidth]{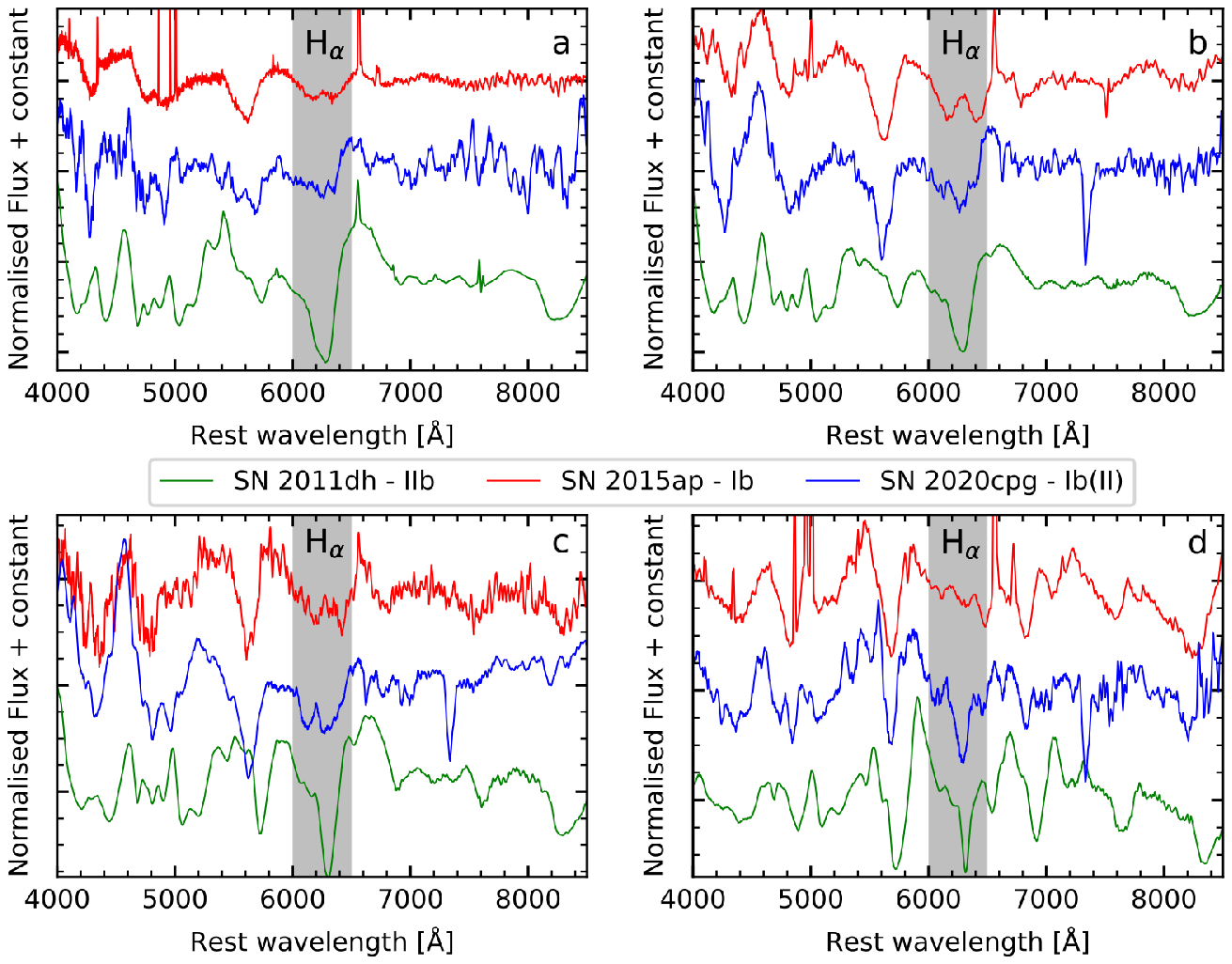}
    \caption{Comparison of SN\,2020cpg (blue) with a characteristic SN Ib (SN\,2015ap, red) and a characteristic SN IIb (SN\,2011dh, green) at several epochs relative to pseudo-bolometric peak. The epochs shown are around a: $-5$ days, b: $+0$ days, c: $+5$ days, d: $+30$ days. The grey shaded area denote the region where the \Ha\ feature should be located if hydrogen is present within the outer envelope of the progenitor. The spectrum used for each of the plots along with the instrument used to obtain them are detailed in Table \ref{Comp_Spec_info} and can all be found on WiseRep \protect\citep{Yaron_2012}.}
    \label{20cpg_Ib/IIb_comp}
\end{figure*}

We compare the spectra of SN\,2020cpg, a H-rich (SN\,2011dh), and a H-poor (SN\,2015ap) SN, within the range $4000 - 9000 \, \si{\angstrom}$ presenting the evolution of the hydrogen features and the line strength relative to standard SNe\,Ib and IIb, see Figure \ref{20cpg_Ib/IIb_comp}. Both SN\,2011dh and SN\,2015ap were close, well observed, SE-SNe allowing for clear comparisons to SN\,2020cpg. This is especially true of SN\,2015ap, which photometrically appears similar to SN\,2020cpg in both shape and luminosity. The epochs chosen were relative to the peak of the pseudo-bolometric light curve so that all SNe were at similar stages in their evolution. The epochs compared are $-5, 0, +5 \text{ and} +30$ days relative to peak luminosity. The grey region in Figure \ref{20cpg_Ib/IIb_comp} highlights the \Ha\ region. It is clear that early on the spectra of SN\,2020cpg are more similar to those of SN\,2015ap, especially in the \HeI\ lines velocity ($\sim 12000$ \kms), and lack of a strong \Ha\ feature. The \HeI\ \lam 6678 feature, which can sometimes blend with the \Ha\ feature is not well defined in SN\,2020cpg at all epochs and can only be clearly seen in plots b and d of Figure \ref{20cpg_Ib/IIb_comp}. As the spectra evolve, the \HeI\ features of SN\,2015ap and SN\,2020cpg deepen in a similar fashion, although the \Ha\ feature of SN\,2020cpg also becomes deeper and more defined. The emergence of the \Ha\ feature results in the spectra of SN\,2020cpg becoming more 2011dh-like and less like those of SN\,2015ap. In the final plot, the spectrum of SN\,2020cpg becomes very similar to that of SN\,2011dh, especially in the \Ha\ region where there is a clear \Ha\ absorption feature that is not present in the SN\,2015ap spectrum. Throughout the emergence of the \Ha\ feature its strength remains weaker than or similar to that of the \HeI\ $\lambda 5876$ peak, compared to the ratio of their strength seen in the SN IIb where the \Ha\ feature dominates throughout the spectra. From the classification scheme of \citet{Prentice_2017} and the strength of the \Ha\ feature relative to the \HeI\ peak, it seems that SN\,2020cpg should be categorised as a Type Ib(II) SN. \par

\section{Discussion}
\label{Discussion}

With a maximum luminosity of $6.03 \pm 0.01 \times 10^{42}$ \ergs, SN\,2020cpg is brighter than the average SE-SNe. Among the SNe we have considered here, only SN\,2009jf and SN\,2015ap have similar luminosities to SN\,2020cpg. This suggests that SN\,2020cpg is brighter than the average SE-SNe. We also compare the maximum luminosity of SN\,2020cpg to the median peak luminosity of SNe\,Ib + Ib(II) and SNe\,IIb + IIb(I) in \citet{2019MNRAS.485.1559P} showing that SN\,2020cpg is located at the brighter end of the luminosity range displayed by H-rich SNe. The rise time of SN\,2020cpg is similar to most other SNe we have looked at, although the pseudo-bolometric light curve of SN\,2020cpg is broader than many of the SN shown in Figure \ref{all_bol+fit}. The high pseudo-bolometric luminosity indicates a large amount of \Nifs. An Arnett-like model fit yielded a total \Nifs\ mass of \mni\,$= 0.27 \pm 0.08$ \msun. From the analysis of several SE-SNe performed by \citet{2019MNRAS.485.1559P}, the Arnett-like model derived mean nickel masses of $<$\mni$>=0.07 \pm 0.03$ \msun\ for SNe\,IIb + IIb(I) and $<$\mni$>=0.09 \pm 0.06$ \msun\ for SNe\,Ib + Ib(II). Therefore, SN\,2020cpg produced roughly triple the mean nickel mass, placing it on the extreme end of SE-SNe. 

Despite the similarity in mean \Nifs\ between SNe\,IIb + IIb(I) and SNe\,Ib + Ib(II), \citet{2019MNRAS.485.1559P} showed that SNe\,Ib + Ib(II) \Nifs\ masses display a bimodal distribution with a high mass region where the \Nifs\ mass of SN\,2020cpg resides. From the distribution of \Nifs\ mass given by \citet{2019MNRAS.485.1559P}, SN\,2020cpg behaves like H-poor SE-SNe.
It should be noted that most neutrino-driven explosion models cannot produce \mni\ greater than $\sim 0.23$ \msun\ \citep{10.1093/mnras/sty3309}, although a study of literature \mni\ values done by \citet{Anderson_2019} found that $\sim 30 \%$ of hydrogen-poor SE-SNe and $\sim 7 \%$ of hydrogen-rich SE-SNe have \Nifs\ masses that are greater $0.23$ \msun. This discrepancy arises from the assumptions of the Arnett-like model, see Section \ref{physical_params}, which result in an overestimation of the \mni. Taking into account this overestimation the \mni\ of SN\,2020cpg is reduced to $\approx 0.16 - 0.19$ \msun, placing SN\,2020cpg's \mni\ within, although close to, the upper limits of neutrino-driven explosion models. However, as we have compared the \mni\ of SN\,2020cpg with other \Nifs\ masses derived by the Arnett-like model and there is uncertainty in the overestimation of the model, we use the value of mass of \Nifs\ derived from the Arnett-like model as the upper limit of \mni\ for SN\,2020cpg.

\subsection{Hydrogen Envelope}

As seen for the spectroscopic evolution of SN\,2020cpg when compared to well observed Type Ib and IIb SNe, there is strong evidence for the presence of a hydrogen envelope surrounding the progenitor of SN\,2020cpg. The separation of the \Ha\ feature into a high velocity and a low velocity component suggests that the hydrogen is located in two distinct regions within the outer envelope of the progenitor star. A thin outer envelope and an inner section where the hydrogen and helium are thoroughly mixed together corresponding to the high and low velocity component respectively. While the two component \Ha\ features are not common among H-rich SE-SNe, it has been observed in other SNe, with SN\,1993J displaying a clear double \Ha\ feature throughout the photospheric phase. The velocity of the high velocity component for SN\,1993J does not seem as large as that for SN\,2020cpg relative to the low velocity component. This suggests that the amount of hydrogen stripped from the progenitor of SN\,2020cpg is greater than that of SN\,1993J prior to the explosion, which is further supported by the weak \Ha\ feature seen in the spectral evolution of SN\,2020cpg. The presence of a weak \Ha\ absorption feature provides evidence that SN\,2020cpg is not a standard Type Ib SNe but rather a Type Ib(II).

\subsection{Model Comparisons}

\begin{table}
    \centering
    \caption{Spectral details for the spectra used in Figure \ref{20cpg_Ib/IIb_comp}, including the date the spectra were obtained and the instrument used to obtain them. All spectra shown in this table and Figure \ref{20cpg_Ib/IIb_comp} are from the sources below;
    1: \protect\citet{2011ApJ...742L..18A}, 2: \protect\citet{2014A&A...562A..17E}, 3: \protect\citet{2019MNRAS.482.1545S}, 4: \protect\citet{2019MNRAS.485.1559P}.}
    \begin{tabular}{ccccc}
        SN & Plot & Date & Instrument & Source \\
        \hline
        \multirow{4}{*}{2011dh} & a & 12/06/2011 & FOS\_1 & 1\\
         & b & 17/06/2011 & ALFOSC & 2\\
         & c & 25/06/2011 & ALFOSC & 2\\
         & d & 14/07/2011 & ALFOSC & 2\\
        \multirow{4}{*}{2015ap} & a & 15/09/2015 & KAST & 3\\
         & b & 20/09/2015 & FLOYDS\_S & 4\\
         & c & 23/09/2015 & FLOYDS\_N & 4\\
         & d & 20/10/2015 & KAST & 3\\
        \multirow{4}{*}{2020cpg} & a & 17/02/2020 & COJ en12 & -\\
         & b & 20/02/2020 & EFOSC2 & -\\
         & c & 23/02/2020 & EFOSC2 & -\\
         & d & 23/03/2020 & EFOSC2 & -\\
         \hline
    \end{tabular}
    \label{Comp_Spec_info}
\end{table}

By comparing the spectra of SN\,2020cpg with model spectra, we can gain insight on the potential elemental composition of the outer layers prior to explosion. \citet{Teffs_2020} calculated a set of synthetic SE-SNe models based on a single mass progenitor, with varying degrees of H/He stripping that produces several Type Ic/Ib/IIb analogue SNe. \citet{Teffs_2020} estimated the energy of a set of well observed Type IIb SNe by comparing synthetic and observed spectra at pre-, near- and post-peak luminosities.

A similar method is applied in this work to SN\,2020cpg. The pre- and near-peak spectra of SN\,2020cpg are very blue and with few strong features. The early synthetic Type IIb-like spectra in \citet{Teffs_2020} are typically redder due to a stronger amount of Fe--group elements mixing, producing strong line blocking in the near UV. As such, the conditions in which the early spectra of SN\,2020cpg are produced are beyond the scope of this comparison and can be explored in future work.

\begin{figure*}
    \begin{subfigure}{\linewidth}
    \centering
    \includegraphics[width=\linewidth, height=10cm]{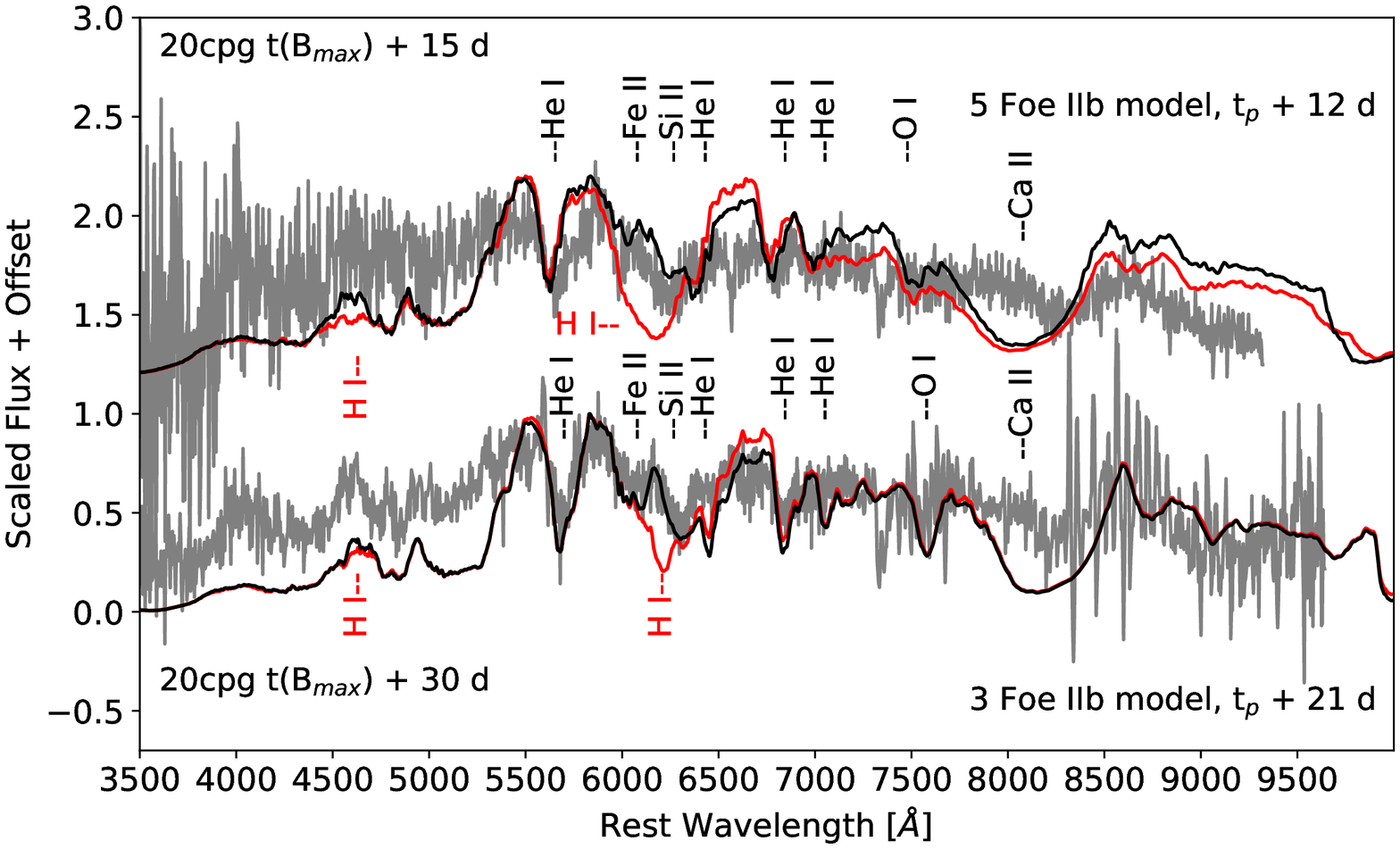}
    \phantomcaption{}\label{IIb_comps}
    \end{subfigure}\\[1ex]
    \begin{subfigure}{\linewidth}
    \centering
    \includegraphics[width=\linewidth, height=10cm]{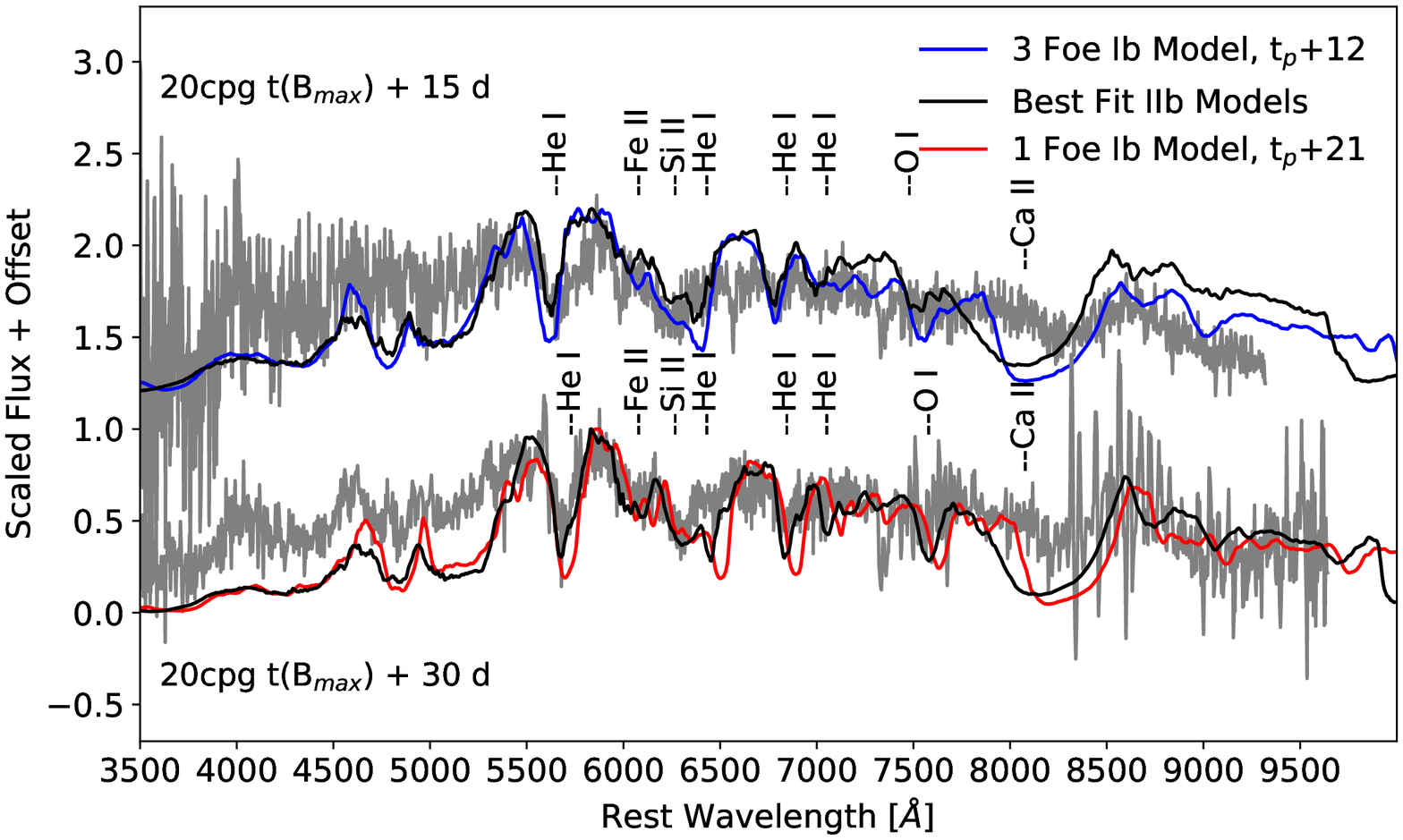}
    \phantomcaption{}\label{Ib_comps}
    \end{subfigure}
    \caption{Top (a): Comparison of SN\,2020cpg spectra with two different energy (3 and 5 foe) Type IIb models. Where $t_\mathrm{p}$ indicates the epoch of bolometric peak. The red model spectra include the non thermal effects of hydrogen while the black model spectra omit the non thermal effects. Both models include the non thermal effects on He.
    Several important contributing elements, in particular those near the \Ha\ line, are shown above their feature within the spectra. Bottom (b): SN\,2020cpg spectra at +15 and +30 days from $B$-band maximum fit with a IIb-like model and with a Ib model with 3 foe of energy at a time of +12 days from $t_\mathrm{p}$ and 1 foe of energy at a time of +21 days respectively. The IIb model has an ejecta mass of $\approx 5.7$ \msun, with $1.3$ \msun\ of helium and $0.1$ \msun\ of hydrogen. }
    \label{Jacob_spec_fits}
\end{figure*}

We first compare two spectra at approximately $+15$ and $+30$ days after $B$-band maximum to the Type IIb model from \citet{Teffs_2020} in Figure \ref{IIb_comps}, where the red spectra include the non--thermal effects on hydrogen and the black do not. These non--thermal effects arise from the interactions with energetic electrons that are created by the scattering of gamma-rays released from the decay chain of \Nifs\ and \Cofs, \citep{1991ApJ...383..308L}. This Type IIb model has an ejecta mass of $\sim$5.7 \msun, with 1.3 \msun\ of helium and 0.1 \msun\ of hydrogen. For the earlier spectrum at the top of Figure \ref{IIb_comps} and when focusing on the \Ha\ and \HeI\ features, we find the best fit to be that of a 5 foe model that does not include the non--thermal effects on H, where 1 foe is $1 \times 10^{51}$ erg. The inclusion of non--thermal hydrogen produces a deep and broader \Ha\ line that is not reflected in the spectrum of SN\,2020cpg. In the second spectrum considered, we instead find that the 3 foe explosion model matches the spectrum of SN\,2020cpg at this late phase. At this lower energy and later phase, the \Ha\ line is more narrow and when the non--thermal effects of hydrogen are not included, the 6000--6500 \AA\ region is well reproduced.

As SN\,2020cpg has been designated both as a Type Ib and a Type IIb, we also compare our helium rich, but hydrogen free, Type Ib models at the same epochs in Figure \ref{Ib_comps}. For this, we also include the "best fit" models from Figure \ref{IIb_comps} that do not consider the non--thermal effects of hydrogen as black lines. For the earlier spectrum, we find that the 3 foe Ib model does a reasonable job of reproducing the 6000--6500 \AA\ region without requiring H, but the \lam6678 \HeI\ is stronger than in the observed spectrum. For the later spectrum, the energy is reduced from the IIb model again to 1 foe and also reproduces this 6000--6500 \AA\ region.

From this, we can infer several properties of the SN\,2020cpg regarding its elemental composition. The assumption that helium has non--thermal effects while H does not is unlikely to be physically viable. However, the mass of hydrogen in the IIb model clearly produces too strong of an \Ha\ line. Not including any hydrogen in the model while maintaining a He--rich outer atmosphere results in strong \lam6678 and \lam7065 \HeI\ lines. The re--emission from the \Ha\ feature reduces the strength of the \lam6678 \HeI\ line while affecting the \lam7065 \HeI\ line less. The best fit Ib models having low energy also suggests the He rich material is confined to lower velocities, such as those below a hydrogen rich shell as seen in the IIb models. We suggests that a lower mass of hydrogen (\mh\ < 0.1 \msun) could result in a weaker \Ha\ feature but still produce enough re-emission to reproduce the 6000--6500 \AA\ region in these late phases. A more detailed model would need to be calculated to derive a stronger estimate on the mass and distribution of H in SN\,2020cpg.

At these two epochs, the photosphere has receded deep into the CO rich region of the ejecta as shown by the presence of the \CaII\ NIR triplet and the \OI\ \lam\ 7771. Early spectra of Type IIb do not show these features as the abundances of these elements are lower in the H/He-rich shells. Both models are shown with (red line) and without (black line) the non thermal effects of hydrogen, but both include these effects on the helium. For the Type IIb-like models at these epochs, the spectra that do not treat the non thermal effects of hydrogen, are better able to reproduce the observed spectral structure between 6000--6500 \AA\ that would typically contain a strong \Ha\ feature. Due to the depth of the photosphere and the lack of a strong early \Ha\ feature, this suggests that the total \Ha\ mass is less that 0.1 \msun, and that the distribution of the hydrogen is further out in the ejecta with respect to the photospheric velocity of the two epochs chosen. \par

\begin{table*}
    \caption{SN parameters obtained from detailed hydrodynamic models and the resulting \mej\ and \ek\ for SN\,2020cpg using the photospheric velocity obtained at the peak of the pseudo-bolometric light curve and the velocity at $t \sim 16$ days post explosion, the time that SN\,2020cpg peaked. The SNe are ordered by their types with the SNe with the least stripped progenitors at the top and most stripped at the bottom. Sources- 1: \protect\citet{Nomoto_1993}, 2: \protect\citet{2002ApJ...572L..61M}, 3: \protect\citet{Sauer_2006}, 4: \protect\citet{Mazzali_2008}, 5: \protect\citet{Tanaka_2009}, 6: \protect\citet{Mazzali_2009}, 7: \protect\citet{Mazzali_2017}
    a; rise time from the explosion to the pseudo-bolometric light curve peak
    b; width of pseudo-bolometric light curve taken from 0.5 mag below peak light
    c; photospheric velocity at epoch of peak light
    d; photospheric velocity at +16 days from explosion date (same epoch as when SN\,2020cpg reached peak light).}
    \centering
    \resizebox{\textwidth}{!}{
    \begin{tabular}[width=\linewidth]{ccccccccccccc}
         & & & & & & & & \multicolumn{2}{c}{2020cpg} & & \multicolumn{2}{c}{2020cpg} \\
        SN & Type & \mej & \ek & Source & $t(\mathrm{L max})_{a}$ & LC width$_\mathrm{b}$ & $v_\mathrm{ph}\mathrm{(max)}_{c}$ & \mej & \ek & $v_\mathrm{ph}\mathrm{(t \sim 16)}_{d}$ & \mej & \ek \\
        & & [\msun] & [$10^{51} $\erg] & & [Days] & [Days] & [\kms] & [\msun] & [$\times 10^{51}$ \erg] & [\kms] & [\msun] & [$\times 10^{51}$ \erg] \\
        \hline
        03bg & IIb-Hyper & $4.8 \pm 0.5$ & $5.0 \pm 1.5$ & $6$ & $25.0 \pm 2.0$ & $34.0 \pm 2.0$ & $7000 \pm 500$ & $3.0 \pm 1.0$ & $11.0 \pm 4.0$ & $9000 \pm 500$ & $2.0 \pm 1.0$ & $4.0 \pm 1.0 $ \\
        93J & IIb & $3.7 \pm 0.5$ & $1.2 \pm 0.1$ & $1$ & $19.0 \pm 2.0$ & $18.0 \pm 2.0$ & $7500 \pm 100$ & $7.0 \pm 2.5$ & $6.0 \pm 1.0$ & $9500 \pm 100$ & $5.0 \pm 2.0$ & $3.0 \pm 1.0 $ \\
        08D & Ib & $7.0 \pm 0.5$ & $6.0 \pm 0.5$ & $4$ & $19.0 \pm 2.0$ & $24.0 \pm 2.0$ & $9500 \pm 500$ & $6.0 \pm 2.0$ & $9.0 \pm 2.0$ & $10000 \pm 500$ & $5.5 \pm 2.0$ & $7.0 \pm 1.0 $ \\
        08D & Ib & $5.3 \pm 1.0$ & $6.0 \pm 0.5$ & $5$ & $19.0 \pm 2.0$ & $24.0 \pm 2.0$ & $9500 \pm 500$ & $4.0 \pm 2.0$ & $9.0 \pm 4.0$ & $10000 \pm 500$ & $4.0 \pm 2.0$ & $7.0 \pm 3.0 $ \\
        04aw & Ic & $4.0 \pm 1.0$ & $4.5 \pm 1.5$ & $7$ & $16.0 \pm 1.0$ & $21.0 \pm 2.0$ & $11500 \pm 500$ & $4.0 \pm 2.0$ & $5.0 \pm 2.0$ & $11500 \pm 500$ & $4.0 \pm 2.0$ & $5.0 \pm 2.0 $ \\
        94I & Ic & $1.2 \pm 0.1$ & $1.0 \pm 0.5$ & $3$ & $12.0 \pm 2.0$ & $11.0 \pm 2.0$ & $10000 \pm 500$ & $4.5 \pm 1.5$ & $6.0 \pm 3.0$ & $9000 \pm 1000$ & $5.0 \pm 2.0$ & $8.0 \pm 5.0 $ \\
        02ap & Ic-BL & $2.4 \pm 1.2$ & $4.0 \pm 0.5$ & $2$ & $13.0 \pm 2.0$ & $18.0 \pm 2.0$ & $12500 \pm 1250$ & $3.0 \pm 2.0$ & $5.0 \pm 1.0$ & $9000 \pm 100$ & $4.0 \pm 3.0$ & $12.0 \pm 3.0 $ \\
        \hline
    \end{tabular}}
    \label{ratio_details}
\end{table*}

The model shown at the top in Figure \ref{IIb_comps} is a 5 foe explosion, with the majority of the 0.1 \msun\ of hydrogen at velocities greater than $\approx$15000 \kms, while the 3 foe explosion in the lower model contains hydrogen at velocities greater than $\approx$12000 \kms. Both models favour both the estimated explosion energy from the Arnett fits in Section \ref{Bol_section} and the suggestion that some hydrogen is at high velocities. The \HeI\ \lam\ 6678 line is relatively too strong for either epoch to match when we do not include non--thermal excitation of the hydrogen. This suggests that a lower mass of hydrogen can still be responsible for some fraction of the 6000-6500 \AA feature, likely coincident with \SiII\, causing a re-emission of flux further redward, reducing the strength of the \HeI\ \lam\ 6678 without affecting the \HeI\ $\lambda\lambda5876, 7065 \text{ and } 7281$ lines. However, for a full picture of how the hydrogen and helium are distributed and how much is present, a detailed stratified model would need to be produced, which is beyond the scope of this work. \par

\subsection{Re-scaled Light curves}
\label{ratio_properties}

As mentioned in Section \ref{physical_params}, the Arnett-like model is limited in its viability to obtain realistic ejecta mass and kinetic energy due to the assumption that the optical opacity is constant throughout the bolometric light curve and that the ejecta are optically thick. The problem with these assumptions is that helium is optically transparent at the temperatures reached surrounding the peak light phase of the light curve. In order to account for the effects of the helium layer on the ejecta mass and kinetic energy a detailed hydrodynamical model is required. However, this would not have been easily done with SN\,2020cpg due to the lack of early time photometry and the low signal to noise ratio for the spectra. In order to estimate the physical parameters for SN\,2020cpg we transform equation \ref{tau_m} to obtain a ratio for the ejecta mass and kinetic energy between SN\,2020cpg and other SE-SNe that have detailed hydrodynamical models:
\begin{ceqn}
\begin{align}
    \frac{E_\mathrm{k1}}{E_\mathrm{k2}} = \frac{\tau_\mathrm{m1}^{2} * v_\mathrm{ph1}^{3} * \kappa_\mathrm{1}^{-1}}{\tau_\mathrm{m2}^{2} * v_\mathrm{ph2}^{3} * \kappa_\mathrm{2}^{-1}},
    \label{E_ratio}
\end{align}
\end{ceqn}
and
\begin{ceqn}
\begin{align}
    \frac{M_\mathrm{ejc1}}{M_\mathrm{ejc2}} = \frac{\tau_\mathrm{m1}^{2} * v_\mathrm{ph1} * \kappa_\mathrm{1}^{-2}}{\tau_\mathrm{m2}^{2} * v_\mathrm{ph2} * \kappa_\mathrm{2}^{-2}}.
    \label{M_ratio}
\end{align}
\end{ceqn}

Where $\tau_\mathrm{m}$ is the diffusion time of the light curve, $v_\mathrm{ph}$ is the photospheric velocity at maximum light and $\kappa$ is the optical opacity of the SN ejecta. Due to the difficulty in determining $\kappa$ we have assumed that it is the same for both SNe. This assumption holds strong for SNe of the same classification type due to the similar elemental structure between the two SNe and becomes weaker as different types of SN are compared to one another. However, as we will be using only SE-SNe to obtain the ejecta mass and kinetic energy of SN\,2020cpg the problem that arises from the use of SNe with different opacities should be minimised. \par

We compare SN\,2020cpg with SN\,1993J \citep{Nomoto_1993}, SN\,1994I \citep{Sauer_2006}, SN\,2002ap \citep{2002ApJ...572L..61M}, SN\,2003bg \citep{Mazzali_2009}, SN\,2004aw \citep{Mazzali_2017} and SN\,2008D \citep{Mazzali_2008, Tanaka_2009}, all SE-SNe that have undergone hydrodynamical modelling. Since several of the above SNe, including SN\,2020cpg, lack early time photometric data, it was not always possible to determine $\tau_\mathrm{m}$. Instead we used the width of the pseudo-bolometric light curve taken from 0.5 mag below peak light as an alternative to $\tau_\mathrm{m}$. Due to the width of the light curve being influenced by both the ejecta mass and kinetic energy, as shown in equation \ref{tau_m}, this allowed for a direct comparison between the widths of the light curves and physical properties of SN\,2020cpg and the modelled SE-SNe. The details on photospheric velocity and light curve widths for each SN along with the \mej\ and \ek\ of SN\,2020cpg given by equation \ref{E_ratio} and equation \ref{M_ratio} are shown in Table \ref{ratio_details}.

We obtain physical parameters for SN\,2020cpg using both the photospheric velocity at pseudo-bolometric peak, $v\mathrm{_{ph}(t=max)}$, for the individual SNe and the photospheric velocity at t = 16 days from the reported explosion date, $v\mathrm{_{ph}(t=16)}$. We use $v\mathrm{_{ph}(t=max)}$ to break the degeneracy between the ejecta mass and the kinetic energy as it would be the velocity of the photosphere when all of the light has diffused through the ejecta. $v\mathrm{_{ph}(t=16)}$ was also used to compare the different SNe at the point when SN\,2020cpg had reached maximum pseudo-bolometric light, allowing a direct comparison between SNe to be made. The values for the physical parameters obtained from comparisons with the hydrodynamical models are higher than those derived from using the Arnett-like model, as expected when comparing the Arnett-like model with hydrodynamical models. The main outlier in Table \ref{ratio_details} is the properties predicted from the SN\,2003bg, a hypernova, which despite having a relatively high kinetic energy possessed a low photospheric velocity at pseudo-bolometric peak, resulting in a low ejecta mass and large kinetic energy. \par

There is a clear trend in the ejecta mass and kinetic energy obtained using the PM13 method which arises from the type of SN that SN\,2020cpg is compared to, with the He-rich SE-SNe resulting in a generally larger values while the values obtained from He-poor SNe are noticeably lower. As SN\,2020cpg is a He-rich SN, we use the physical parameters obtained from the Type Ib and IIb SNe to determine the values of the ejected mass, \mej\ $\sim 5.5 \pm 2.0$ \msun, and kinetic energy, \ek\ $\sim 9.0 \pm 3.0 \times 10^{51}$ \erg, for SN\,2020cpg. The value obtained using $v\mathrm{_{ph}(t=16)}$ for the ejecta mass was $\sim 4.0 \pm 1.5$\,\msun\ and a kinetic energy of $\sim 5.0 \pm 2.0 \times 10^{51}$ \erg. It should be noted that the PM13 model is limited in scope and should not be expected to predict values of both $M_\mathrm{ejc}$ and $E_\mathrm{k}$ to a precision greater than $0.5$ \msun\ and $1.0 \times 10^{51}$ \erg\ respectively. The values produced using $v\mathrm{_{ph}(t=max)}$ converge on the physical parameters with an average standard deviation of $1.85$ while the $v\mathrm{_{ph}(t=16)}$ has an average standard deviation of $1.89$. This suggests that using the photospheric velocity at pseudo-bolometric peak for each SN converge on a value better than the photospheric velocity at SN\,2020cpg pseudo-bolometric peak. The values for the ejecta mass and kinetic energy produced by the PM13 method are much higher than those predicted by the Arnett-like model, as expected due to the contribution of the helium envelope and the effect of having similar optical opacities. However, the ejecta mass derived from the PM13 method matches the value obtained by comparing the spectra of SN\,2020cpg with the spectral models of \citet{Teffs_2020}, although the kinetic energy given by the modelling is lower than that predicted using the method from PM13. \par

The ejecta mass given by the spectral modelling and comparison with modelled SE-SNe has a value roughly double that given for Ib + Ib(II) and IIb + IIb(I) by \citet{2019MNRAS.485.1559P} which take a mean value of $2.2 \pm 0.9$ \msun\ and $2.7 \pm 1.0$ \msun\ respectively. This places SN\,2020cpg in the higher mass range of SE-SNe with only one H-rich and two H-poor SE-SNe having similar ejecta mass. The ejecta mass predicted by the Arnett-like model is closer to the mean values given by \citet{Prentice_2017} although still greater than the median, showing that by all standards SN\,2020cpg was a more massive event than the typical SE-SNe. The lower ejecta mass estimated by the Arnett-like model is expected, as this has been seen in several SNe such as SN\,2008D which was estimated to have an ejecta mass of $2.9^{+ 1.0}_{-0.6}$\,\msun\ from an Arnett-like approach \citep{Lyman_2016} and $\sim 5 - 7$ \msun\ from hydrodynamic modelling \citep{Mazzali_2008, Tanaka_2009}. When compared to the ejecta masses of the H-rich SE-SNe, SN\,2020cpg lies in the region that has been associated with an extended progenitor. As the ejecta mass obtained using the comparative method of PM13 and the spectral modelling of \citet{Teffs_2020} are in close agreement, we take the PM13 method as a valid replacement for the Arnett-like model to obtain the ejecta mass when dealing with SE-SNe. The PM13 method will also improve in the future as more SE-SNe undergo hydrodynamical modelling. \par

From the PM13 method the derived kinetic energy takes a value of $\sim 9.0 \pm 3.0 \times 10^{51}$ \erg\ which is greater than both the spectral modelling and the Arnett-like model. This kinetic energy place SN\,2020cpg on the border of the Hypernovae, which are thought to have kinetic energies on the order of $10^{52}$ \erg. The kinetic energy derived from spectral modelling tends towards a lower kinetic energy than the PM13 method, however, larger than the kinetic energy estimated by the Arnett-like model. The Arnett-like model derived a kinetic energy of $\sim 2.9 \pm 0.9 \times 10^{51}$ \erg, which is similar to the kinetic energy suggested by the spectral modelling. However, given the high \Nifs\ mass the kinetic energy derived from the Arnett-like model is unlikely to be enough to synthesis the required amount of nickel.

From the derived ejecta mass, under the assumption that the progenitor did not collapse into a black hole but holds a 1.4 \msun\ neutron star, the progenitors core mass can be assumed to be $M_\mathrm{ejc} + M_\mathrm{NS} - M_\mathrm{outer envelope} = M_\mathrm{COcore} \approx 6.0 \pm 2.0$ \msun. Here we assumed that the mass of the outer envelope was $\sim 1.5$ \msun. This core mass is just higher than the majority of SE-SNe investigated in \citet{Prentice_2017} which takes a mean value $< 5$ \msun. A core mass of $\sim 6.0 \pm 2.0$ \msun\ is thought to originate from a progenitor with an initial mass of $18 - 25$ \msun, \citep{2016ApJ...821...38S}.
suggests that the progenitor of SN\,2020cpg would have had a high mass prior to explosion, within the range of $18 - 25$ \msun.

\begin{table}
    \caption{Opacities derived from the SN\,2020cpg opacity. There seems to be a trend with the He-rich SNe having a lower opacity than the He-poor SNe.}
    \centering
    \begin{tabular}[width=\columnwidth]{ccc}
    SN & Type & Opacity $\mathrm{[cm^{2}g^{-1}]}$ \\
    \hline
    03bg & IIb-Hyper & $0.27\pm0.19$ \\
    93J & IIb & $0.06 \pm 0.04$ \\
    08D & Ib & $0.08 \pm 0.06$ \\
    08D & Ib & $0.13 \pm 0.09$ \\
    04aw & Ic & $0.13 \pm 0.10$ \\
    94I & Ic & $0.10 \pm 0.08$ \\
    02ap & Ic-BL & $0.19 \pm 0.16$ \\
    \hline
    \end{tabular}
    \label{opacities}
\end{table}

As mentioned earlier with the Arnett-like model, the opacity for both SN\,2020cpg and the comparison SN is neither constant nor the same. To this end we used equation \ref{tau_m} to obtain an opacity for SN\,2020cpg, which had a value of $\kappa_\mathrm{opt} = 0.10 \pm 0.04$ cm$^{2}$~g$^{-1}$. Using the opacity for SN\,2020cpg we then obtained the opacities of all the SE-SNe we compared with SN\,2020cpg, which are shown in Table \ref{opacities}. As expected the He-rich SNe tend to have a lower opacity than the He-poor SNe, due to the fact that the helium present within the ejecta is virtually transparent to the optical photons. The opacity for SN\,2008D has two values due to the different ejecta masses that we used. By looking at the opacities determined using the above method it is clear that a single time-independent value of the opacity should not be used for all types of SE-SNe, as done with the Arnett-like model discussed in Section \ref{physical_params}.

\section{Summary and Conclusions}
\label{Sum+Con}
The study of SN\,2020cpg and the discovery of the weak hydrogen features within the otherwise SN Ib-like spectra shows formation channels between SNe\,Ib and IIb are not as rigid as previously thought. From the coverage of SN\,2020cpg we were able to compare the evolution of SN\,2020cpg with several other SE-SNe. Photometrically, SN\,2020cpg looks very similar to the Type Ib SN\,2009jf in peak luminosity, although the light curve of SN\,2020cpg is slightly broader compared to SN\,2009jf. Spectroscopically SN\,2020cpg initially looked similar to the Type Ib SN, such as SN\,2015ap, with the main difference being the presence of the weak \Ha\ feature within the spectra of SN\,2020cpg. As the spectra evolve, the \Ha\ feature becomes more dominant until it rivals the \HeI\ $\lambda 5876$ feature in strength, making SN\,2020cpg resemble more that of a Type IIb SN, such as SN\,2011dh. Due to the weak \Ha\ feature that is shown within the spectra of SN\,2020cpg we believe that it was a Type Ib(II) SN. As the \Ha\ feature grows in strength from the initial spectrum, we suggest that the hydrogen may have existed in a thin envelope as well as mixed into the outer layers of the helium shell prior to the explosion, which became more dominant as the photosphere receded through the mixed hydrogen/helium layer.\par

SN\,2020cpg exploded producing an estimated Nickel mass of $\sim 0.3\pm 0.1$ \msun\, and from comparisons with hydrodynamic models of well studied He-rich SE-SNe an ejecta mass of $\sim 5.5 \pm 2.0$ \msun\ and a kinetic energy of $\sim 9.0 \pm 3.0 \times 10^{51}$ erg. From spectral modelling the amount of helium expected within the ejecta is $1.3$ \msun\ with a further $0.1$ \msun\ of hydrogen contained within the outer envelope with a large majority of it existing above a velocity of $\approx 15000$ \kms. From this modelling and the assumption that a neutron star remnant was formed, SN\,2020cpg would have had a core mass of $M_\mathrm{core} = 6.0 \pm 2.0$ \msun\ which corresponds to a progenitor star with a initial mass of \mzams\ $\sim 18 - 25$ \msun. Due to the distance to the host galaxy and the position of SN\,2020cpg within the host galaxy, it is unlikely that there are any pre-explosion images of high enough quality to allow for the progenitor of SN\,2020cpg to be determined. Further modelling of SN\,2020cpg may give evidence for the progenitor however that is beyond the scope of this paper.

The use of the PM13 model provides an alternative approach to the Arnett-like model in determining the ejecta mass and kinetic energy of new SE-SNe. The PM13 method accounts for the effects of the helium layer and the time dependency of the optical opacity, both of which are ignored in the Arnett-like approach. The PM13 model produces ejecta masses and kinetic energies that resemble those derived from comparison of optical spectra with spectral models, where as the Arnett-like approach seems to underestimate these values. Unlike the Arnett-like model, when used on SE-SNe the PM13 model requires several SNe of the same classification to constrain the ejecta mass and kinetic energy. This can lead to some outliers, like hypernovae, distorting the results. However as more SE-SNe undergo hydrodynamical modelling the constraining power of the PM13 model increase and the effect the outliers have is reduced. 

\section*{Acknowledgements}
SJP is supported by H2020 ERC grant no.~758638. JJT is funded by the consolidated STFC grant no. R27610. This paper is based in part on observations collected at the European Southern Observatory under ESO programme 1103.D-0328(J). TWC acknowledges the EU Funding under Marie Sk\l{}odowska-Curie grant H2020-MSCA-IF-2018-842471. LG was funded by the European Union's Horizon 2020 research and innovation programme under the Marie Sk\l{}odowska-Curie grant agreement No. 839090. This work has been partially supported by the Spanish grant PGC2018-095317-B-C21 within the European Funds for Regional Development (FEDER). MG is supported by the Polish NCN MAESTRO grant 2014/14/A/ST9/00121. TMB was funded by the CONICYT PFCHA / DOCTORADOBECAS CHILE/2017-72180113. MN is supported by a Royal Astronomical Society Research Fellowship. This work makes use of observations obtained by the Las Cumbres Observatory global telescope network. The LCO team is supported by NSF grants AST-1911225 and AST-1911151. Based in part on observations made with the Liverpool Telescope operated on the island of La Palma by Liverpool John Moores University in the Spanish Observatorio del Roque de los Muchachos of the Institutode Astrofisica de Canarias with financial support from the UK Science and Tech-nology Facilities Council. The data presented here were obtained in part with ALFOSC, which is provided by the Instituto de As-trofisica de Andalucia (IAA) under a joint agreement with the University of Copenhagen and NOTSA, with observation having been made with the Nordic Optical Telescope, operated at the Observatorio del Roque de los Muchachos, La Palma, Spain, of the Instituto de Astrofisica de Canarias. This work has made use of data from the Asteroid Terrestrial-impact Last Alert System (ATLAS) project. ATLAS is primarily funded to search for near earth asteroids through NASA grants NN12AR55G, 80NSSC18K0284, and 80NSSC18K1575; byproducts of the NEO search include images and catalogues from the survey area. The ATLAS science products have been made possible through the contributions of the University of Hawaii Institute for Astronomy, the Queen’s University Belfast, and the Space Telescope Science Institute. Based on observations collected at the European Organisation for Astronomical Research in the Southern Hemisphere, Chile, as part of ePESSTO+ (the advanced Public ESO Spectroscopic Survey for Transient Objects Survey). ePESSTO+ observations were obtained under ESO program ID 1103.D-0328 (PI: Inserra). LCO data have been obtained via OPTICON proposals (IDs: SUPA2020B-002 SUPA2020A-001 OPTICON 20A/015 and OPTICON 20B/003). The OPTICON project has received funding from the European Union's Horizon 2020 research and innovation programme under grant agreement No 730890.

\section*{Data Availability}
Data will be made available on the Weizmann Interactive Supernova Data Repository (WISeREP) at https://wiserep.weizmann.ac.il/.



\bibliographystyle{mnras}
\bibliography{Reference} 



\appendix
\section{Photometric Observations}
\label{Appendix}
\onecolumn
\begin{table*}
    \centering
        \caption{Apparent $BgVri$ LCO Photomety of SN\,2020cpg, no k-correction or extinction correction applied.}
    \begin{tabular}[width=\columnwidth]{cccccccccc}
$MJD_B$ & $B(err)$ & MJD$_{g\text{'}}$ & $g\text{'}(err)$ & MJD$_V$ & $V(err)$ & MJD$_{r\text{'}}$ & $r\text{'}(err)$ & MJD$_{i\text{'}}$ & $i\text{'}(err)$ \\
 & [mag] & & [mag] & & [mag] & & [mag] & & [mag] \\
\hline
58900.362 & 18.39(0.02) & 58894.544 & 18.55(0.09) & 58900.367 & 18.25(0.02) & 58894.501 & 18.49(0.08) & 58900.380 & 18.41(0.02) \\
58900.364 & 18.37(0.02) & 58900.371 & 18.05(0.01) & 58900.369 & 18.28(0.02) & 58900.376 & 18.35(0.02) & 58900.382 & 18.40(0.02) \\
58902.316 & 18.35(0.02) & 58900.374 & 18.20(0.01) & 58902.322 & 18.20(0.02) & 58900.378 & 18.24(0.02) & 58902.335 & 18.24(0.02) \\
58902.319 & 18.36(0.02) & 58902.326 & 18.08(0.01) & 58902.324 & 18.18(0.02) & 58902.331 & 18.16(0.02) & 58902.336 & 18.22(0.02) \\
58903.337 & 18.49(0.02) & 58902.328 & 18.06(0.01) & 58903.343 & 18.10(0.02) & 58902.333 & 18.14(0.02) & 58903.355 & 18.20(0.02) \\
58903.340 & 18.45(0.02) & 58903.346 & 18.06(0.01) & 58903.345 & 18.08(0.02) & 58903.352 & 18.16(0.02) & 58903.357 & 18.17(0.02) \\
58905.101 & 18.35(0.02) & 58903.349 & 18.07(0.01) & 58905.238 & 18.10(0.02) & 58903.354 & 18.13(0.02) & 58905.251 & 18.17(0.02) \\
58905.233 & 18.39(0.02) & 58905.242 & 18.05(0.01) & 58905.240 & 18.11(0.02) & 58905.247 & 18.07(0.02) & 58905.253 & 18.18(0.02) \\
58905.236 & 18.41(0.02) & 58905.245 & 18.06(0.01) & 58906.256 & 18.24(0.02) & 58905.249 & 18.09(0.02) & 58906.269 & 18.06(0.02) \\
58906.251 & 18.47(0.02) & 58906.260 & 18.10(0.01) & 58906.258 & 18.25(0.02) & 58906.265 & 18.11(0.02) & 58906.271 & 18.04(0.02) \\
58906.254 & 18.56(0.02) & 58906.263 & 18.09(0.01) & 58907.277 & 18.21(0.02) & 58906.267 & 18.08(0.02) & 58907.290 & 18.12(0.02) \\
58907.272 & 18.59(0.02) & 58907.281 & 18.14(0.01) & 58907.279 & 18.24(0.02) & 58907.286 & 18.02(0.02) & 58907.291 & 18.06(0.02) \\
58907.274 & 18.58(0.02) & 58907.283 & 18.16(0.01) & 58909.253 & 18.35(0.02) & 58907.288 & 18.05(0.02) & 58909.265 & 18.08(0.02) \\
58909.248 & 18.63(0.02) & 58909.256 & 18.16(0.02) & 58909.255 & 18.37(0.02) & 58909.262 & 18.01(0.02) & 58909.267 & 18.03(0.02) \\
58909.250 & 18.73(0.02) & 58909.259 & 18.17(0.02) & 58910.336 & 18.09(0.02) & 58909.264 & 18.04(0.02) & 58910.349 & 18.01(0.02) \\
58910.331 & 18.48(0.02) & 58910.340 & 18.29(0.02) & 58910.338 & 18.06(0.02) & 58910.345 & 18.07(0.02) & 58910.350 & 18.06(0.02) \\
58910.333 & 18.52(0.02) & 58910.342 & 18.23(0.02) & 58912.105 & 18.18(0.02) & 58910.347 & 18.10(0.02) & 58912.118 & 18.05(0.04) \\
58912.100 & 18.77(0.02) & 58912.109 & 18.37(0.02) & 58912.107 & 18.34(0.02) & 58912.114 & 18.11(0.02) & 58912.119 & 18.06(0.04) \\
58912.102 & 18.78(0.02) & 58912.111 & 18.39(0.02) & 58914.389 & 18.45(0.02) & 58912.116 & 18.09(0.02) & 58914.401 & 18.09(0.03) \\
58914.383 & 19.01(0.02) & 58914.392 & 18.63(0.02) & 58914.390 & 18.44(0.02) & 58914.397 & 18.17(0.02) & 58914.403 & 18.14(0.03) \\
58914.386 & 19.10(0.02) & 58914.395 & 18.55(0.02) & 58916.355 & 18.76(0.02) & 58914.399 & 18.18(0.02) & 58916.368 & 18.19(0.02) \\
58916.350 & 19.38(0.04) & 58916.359 & 18.80(0.02) & 58916.357 & 18.70(0.02) & 58916.364 & 18.20(0.02) & 58916.369 & 18.22(0.02) \\
58916.352 & 19.37(0.04) & 58916.361 & 18.80(0.02) & 58917.323 & 18.48(0.02) & 58916.366 & 18.28(0.02) & 58917.347 & 18.37(0.05) \\
58917.313 & 19.31(0.05) & 58917.331 & 18.89(0.02) & 58917.327 & 18.56(0.02) & 58917.341 & 18.28(0.02) & 58917.349 & 18.49(0.05) \\
58917.318 & 19.24(0.05) & 58917.336 & 18.98(0.02) & 58920.620 & 18.62(0.06) & 58917.344 & 18.34(0.02) & 58920.643 & 18.46(0.05) \\
58920.610 & 19.47(0.09) & 58920.627 & 19.24(0.02) & 58920.623 & 18.71(0.06) & 58920.638 & 18.54(0.02) & 58920.646 & 18.40(0.05) \\
58920.615 & 19.47(0.09) & 58920.632 & 19.14(0.02) & 58924.184 & 18.87(0.04) & 58920.640 & 18.50(0.02) & 58924.208 & 18.40(0.03) \\
58924.174 & 19.94(0.06) & 58924.192 & 19.45(0.02) & 58924.188 & 18.85(0.04) & 58924.202 & 18.59(0.02) & 58924.210 & 18.40(0.03) \\
58924.179 & 19.90(0.06) & 58924.197 & 19.57(0.02) & 58927.094 & 19.09(0.04) & 58924.205 & 18.60(0.02) & 58927.118 & 18.59(0.02) \\
58927.084 & 20.25(0.06) & 58927.102 & 19.75(0.02) & 58927.098 & 19.00(0.04) & 58927.112 & 18.81(0.02) & 58927.120 & 18.62(0.02) \\
58927.089 & 20.19(0.06) & 58927.107 & 19.67(0.02) & 58931.054 & 19.37(0.04) & 58927.115 & 18.85(0.02) & 58931.080 & 18.81(0.01) \\
58931.044 & 20.65(0.05) & 58931.062 & 20.11(0.02) & 58931.058 & 19.52(0.04) & 58931.072 & 19.09(0.02) & 58931.084 & 18.78(0.01) \\
58931.049 & 20.42(0.05) & 58931.067 & 20.05(0.02) & 58949.708 & 19.92(0.12) & 58931.076 & 19.15(0.02) & 58939.657 & 19.12(0.01) \\
- & - & - & - & 58951.685 & 19.99(0.08) & 58939.653 & 19.34(0.02) & 58951.692 & 19.53(0.03) \\
- & - & - & - & 58959.235 & 20.11(0.05) & 58951.689 & 19.66(0.04) & 58959.243 & 19.54(0.04) \\
- & - & - & - & 58974.207 & 20.20(0.15) & 58959.239 & 19.77(0.03) & 58974.215 & 19.76(0.11) \\
- & - & - & - & 58982.144 & 20.48(0.15) & 58974.211 & 20.01(0.10) & 58982.151 & 19.88(0.05) \\
- & - & - & - & 58985.538 & 20.53(0.07) & 58982.147 & 20.03(0.04) & 58985.546 & 19.91(0.05) \\
- & - & - & - & 58993.431 & 20.36(0.09) & 58985.542 & 20.07(0.03) & 58993.439 & 20.14(0.09) \\
- & - & - & - & 59000.823 & 20.67(0.14) & 58993.435 & 20.17(0.11) & 59000.831 & 20.43(0.15) \\
- & - & - & - & 59008.891 & 20.54(0.15) & 59000.827 & 20.21(0.10) & 59008.898 & 20.17(0.06) \\
- & - & - & - & - & - & 59008.894 & 20.40(0.12) & - & - \\
\hline
    \end{tabular}
    \label{All_photometry}
\end{table*}
\newpage
\begin{table}
    \centering
     \caption{Apparent c+o band ATLAS photometry for SN\,2020cpg. Photometry has not been corrected for either extinction or k-correction.}
    \begin{tabular}[height=\paperhigher]{cccccc}
MJD$_{c}$ & $c(err)$ & MJD$_{o}$ & $o(err)$ & MJD$_{o}$ & $o(err)$ \\
 & [mag] & & [mag] & & [mag] \\
 \hline
58903.464 & 18.10(0.06) & 58901.489 & 18.33(0.08) & 58953.488 & 19.32(0.17) \\
58903.499 & 18.10(0.05) & 58901.493 & 18.23(0.07) & 58957.400 & 19.80(0.30) \\
58903.503 & 18.19(0.06) & 58901.500 & 18.24(0.06) & 58957.412 & 19.59(0.20) \\
58903.512 & 18.18(0.05) & 58901.511 & 18.22(0.06) & 58957.415 & 19.87(0.27) \\
58911.503 & 18.13(0.05) & 58905.562 & 18.07(0.05) & 58961.441 & 19.38(0.16) \\
58931.502 & 19.27(0.13) & 58905.565 & 18.09(0.05) & 58961.444 & 19.78(0.22) \\
58931.522 & 19.59(0.17) & 58905.573 & 18.06(0.05) & 58961.465 & 19.67(0.21) \\
58931.530 & 19.45(0.15) & 58905.583 & 18.10(0.05) & 58965.446 & 19.26(0.20) \\
58931.539 & 19.46(0.16) & 58913.454 & 18.20(0.07) & 58965.457 & 19.73(0.29) \\
58935.583 & 19.78(0.25) & 58913.458 & 18.10(0.06) & 58965.491 & 19.55(0.21) \\
58935.586 & 19.55(0.21) & 58913.463 & 18.24(0.07) & 58969.404 & 20.12(0.30) \\
58959.426 & 20.14(0.30) & 58913.477 & 18.19(0.07) & 58969.417 & 19.83(0.23) \\
58959.430 & 19.66(0.19) & 58917.449 & 18.25(0.18) & 58969.421 & 20.02(0.27) \\
58959.438 & 20.18(0.28) & 58917.457 & 18.32(0.19) & 58969.428 & 19.75(0.22) \\
58959.455 & 20.05(0.27) & 58917.467 & 18.78(0.27) & 58971.423 & 19.68(0.26) \\
58967.415 & 19.96(0.23) & 58925.533 & 18.70(0.21) & 58971.439 & 19.54(0.25) \\
58967.425 & 20.10(0.26) & 58925.538 & 18.65(0.17) & 58977.446 & 19.21(0.28) \\
58967.460 & 20.10(0.28) & 58933.537 & 19.36(0.18) & 58981.387 & 19.63(0.20) \\
58982.391 & 20.06(0.25) & 58933.543 & 19.02(0.13) & 58981.404 & 20.07(0.30) \\
58987.399 & 20.12(0.27) & 58933.547 & 19.14(0.15) & 58981.415 & 19.76(0.26) \\
- & - & 58933.559 & 18.86(0.12) & 58985.366 & 19.94(0.23) \\
- & - & 58937.496 & 19.27(0.13) & 58985.380 & 20.12(0.28) \\
- & - & 58937.498 & 19.32(0.12) & 58989.354 & 19.76(0.18) \\
- & - & 58937.508 & 18.99(0.29) & 58989.358 & 20.08(0.25) \\
- & - & 58937.522 & 19.47(0.17) & 58989.393 & 20.06(0.24) \\
- & - & 58941.433 & 18.80(0.13) & 58997.324 & 20.07(0.28) \\
- & - & 58941.445 & 18.91(0.15) & 58997.331 & 20.13(0.28) \\
- & - & 58941.448 & 19.50(0.24) & 58997.335 & 20.06(0.30) \\
- & - & 58941.463 & 19.18(0.27) & 58999.352 & 19.83(0.28) \\
- & - & 58943.479 & 19.44(0.25) & 58999.355 & 19.77(0.29) \\
- & - & 58943.484 & 19.47(0.30) & 59006.346 & 19.44(0.30) \\
- & - & 58949.477 & 19.20(0.26) & 59013.351 & 20.26(0.30) \\
- & - & 58949.482 & 19.30(0.30) & 59013.357 & 20.25(0.29) \\
- & - & 58951.461 & 18.89(0.28) & 59013.367 & 20.23(0.28) \\
- & - & 58951.469 & 19.57(0.29) & 59021.328 & 20.08(0.27) \\
- & - & 58951.482 & 19.17(0.29) & 59021.346 & 20.04(0.27) \\
- & - & 58953.467 & 19.28(0.19) & 59025.321 & 20.26(0.30) \\
- & - & 58953.474 & 19.92(0.28) & 59029.318 & 19.44(0.30) \\
- & - & 58953.478 & 19.46(0.19) & 59037.297 & 19.95(0.29) \\
\hline
    \end{tabular}
    \label{ATLS_photometry}
\end{table}



\bsp	
\label{lastpage}
\end{document}